\documentclass[sigplan,10pt]{acmart}

\usepackage{booktabs}   % professional tables
\usepackage{graphicx}   % figures (figures/ is on the graphics path, see below)
\usepackage{subcaption} % subfigures
\usepackage{listings}   % code / config listings
\usepackage{tikz}       % timing / schematic diagrams
\usepackage{float}      % [H] to force a float exactly in place
\usetikzlibrary{patterns}
\usetikzlibrary{arrows.meta} % arrowheads for the differentiated-service timeline

\graphicspath{{figures/}{../figures/}}

\definecolor{dpgreen}{HTML}{DDE8CB}
\definecolor{zbred}{HTML}{FCB4AF}

\renewcommand\footnotetextcopyrightpermission[1]{} % suppress copyright-permission footnote
\begin{document}
\pagestyle{plain}                                   % plain running pages (with folios)
\raggedbottom                                       % don't stretch columns to flush bottom
                                                    % (avoids large gaps under short floats; revert to
                                                    % \flushbottom for camera-ready if required)

%% ---------------------------------------------------------------------------
%% Title and authors
%% ---------------------------------------------------------------------------
% \texorpdfstring keeps the typeset line break but gives hyperref a clean
% (break-free) string for the PDF metadata, avoiding a spurious warning.
\title{USB Differentiated Service for Networked Cyber-Physical Systems}
% \subtitle{Working draft — ATC 2026}

\author{Anton Njavro}
\affiliation{%
  \institution{Boston University}
  \city{Boston}
  \country{USA}}
\email{njavro@bu.edu}

\author{Rafiuddin Syed}
\affiliation{%
  \institution{Drako Motors}
  \city{San Jose}
  \country{USA}}
\email{rafiuddin.syed@drakomotors.com}

\author{Shriram Raja}
\affiliation{%
  \institution{Boston University}
  \city{Boston}
  \country{USA}}
\email{shriramr@bu.edu}

\author{Zhiyuan Ruan}
\affiliation{%
  \institution{Boston University}
  \city{Boston}
  \country{USA}}
\email{zruan@bu.edu}

\author{Richard West}
\affiliation{%
  \institution{Boston University}
  \city{Boston}
  \country{USA}}
\email{richwest@bu.edu}

% \author{Second Author}
% \affiliation{%
%   \institution{TODO}
%   \city{TODO}
%   \country{TODO}}
% \email{todo@example.com}

\renewcommand{\shortauthors}{Njavro et al.}

%% ---------------------------------------------------------------------------
%% Abstract
%% ---------------------------------------------------------------------------
\begin{abstract}
  Cyber-physical systems are increasingly complex, requiring high throughput, low-latency data processing across multiple cores and compute devices. Additional functional safety challenges dictate the need for high system availability in the presence of failures. Together, these requirements are addressed using a network of compute nodes across which data processing, control and functional safety tasks are distributed. A key challenge is how to address the service requirements of independent data flows, spanning a set of hosts and I/O devices.
%
%  This paper is motivated by earlier work that shows how to support
%  interrupt-based differentiated service for I/O devices. Our
%  contribution is to extend that approach to support differentiated
%  service across a network of compute nodes using USB.

  In this paper, we show how to develop connectivity between hosts using a custom xDCI gadget driver that uses USB interrupters to differentiate the processing of separate data flows. Our solution is the first to handle both isochronous and bulk communication at USB 3.0 transfer rates, with predictable bus latencies down to the granularity of a 125 microsecond microframe, which is sufficient for many CPS control operations. We present an end-to-end solution that couples the predictable execution of user-level tasks with interrupt processing and data communication, so that multiple virtual connections are able to share a common bus. This work is not only applicable to host networking but also device I/O, which makes it arguably desirable for CPS scenarios requiring short-range communication, such as in vehicle control area networks or robotics.
\end{abstract}

%% CCS concepts and keywords (optional but recommended for ACM formats).
%% Generate the \ccsdesc lines from https://dl.acm.org/ccs and paste here.
% \begin{CCSXML}
% \end{CCSXML}
% \ccsdesc[500]{Networks~Network performance evaluation}


\maketitle

%% ---------------------------------------------------------------------------
%% Body — each section lives in its own file under src/sections/
%% ---------------------------------------------------------------------------
\section{Introduction}
\label{sec:intro}

% Motivating comparison table. Reference it with Table~\ref{tab:interconnects}
% once the intro prose is written. Spans both columns (table*) for readability.
\begin{table*}[!htb]
  \centering
  \caption{Common interconnect technologies for cyber-physical systems and their
    principal trade-offs.}
  \label{tab:interconnects}
  \begin{tabular}{@{}llll@{}}
    \toprule
    \textbf{Interconnect} & \textbf{Pros}          & \textbf{Cons}              & \textbf{Use case} \\
    \midrule
    \textbf{CAN} & Ubiquitous in CPS      & Low throughput             & Automotive, embedded \\
    \textbf{USB} & High bandwidth support & No peer networking support & Consumer I/O \\
    \textbf{TSN} & Real-time networking   & No I/O support             & Automotive, factory automation \\
    \bottomrule
  \end{tabular}
\end{table*}
Cyber-physical systems (CPS) couple computation to the physical world. Data processing pipelines span from sensors to actuators, and must keep pace with the physical processes under control.
Modern CPS workloads push
this pipeline beyond a single machine. High-rate sensors demand more
processing than one node provides. Functional safety demands availability in
the presence of failures, which requires redundant nodes. The result is a
network of compute nodes among which sensing, control, and safety tasks are
distributed. A sensing-to-actuation pipeline therefore interacts with both
I/O devices and other hosts, and it needs 
% communication with low latency and high bandwidth
low latency, high bandwidth communication 
at every stage.

Three interconnects dominate I/O and networking in cyber-physical systems.
Table~\ref{tab:interconnects} summarizes their trade-offs. CAN is ubiquitous in
CPS and well studied from a real-time standpoint. Its throughput, however, falls
short of modern I/O. Even CAN~XL, its fastest variant, is limited to data
rates of 10--20\,Mbit/s~\cite{cia-canxl}. USB offers high throughput and dominates the I/O domain, but it was not originally designed to support host-to-host networking. Consequently, it has thus far mostly been resigned to consumer electronics. TSN provides real-time networking and is a comparatively
recent technology, finding adoption in automotive and embedded systems. However, it is not commonly used for I/O device communication, so task pipelines spanning sensors, host processing nodes and actuators must bridge to other buses.

Recent work~\cite{njavro2025real} showed how to use USB to bridge host
networking and device I/O.
%  using USB for all bus links. 
That work was limited to using USB 2.0 bulk communication without differentiated service support for separate pipes sharing the same bus. This work extends the prior art to include differentiated services, based on multiple USB interrupters, for host-to-host and device-host communication. We build on USB, which already pairs high
throughput with broad I/O adoption, and add the peer-to-peer networking it lacks.

Today, host networking and device I/O is split across two fabrics. For example, USB connects devices to a host, while Ethernet or CAN connects hosts to one another. Each fabric brings its
own software stack, and traffic that crosses from one to the other must be
bridged between them. The split adds cost, wiring, and software complexity. A
single bus that serves both roles removes the bridging problem and simplifies
the stack. USB is the natural candidate, already being the bus of choice for
high-rate CPS sensors such as cameras and LiDARs. Its signaling rate has
grown to 5~Gbps at SuperSpeed and 40~Gbps in USB~4, comparable to dedicated
network fabrics. We target short-range communication, as found in robots and
vehicles, where a few meters of cable connect every node.

Two challenges stand in the way. The first is handling asymmetry, resulting from the master-slave nature of USB. A host operates as the bus master, ensuring a device only sends data when it initiates a transfer request. Nothing in this model lets two hosts communicate as peers. The second challenge is guaranteeing differentiated service. A consolidated CPS host carries independent
flows with different requirements. For example, a camera stream needs sustained bandwidth, while a control flow needs bounded delay, and a file transfer require lossless communication. Commodity
USB stacks funnel the completions of all flows through one path, so the
service each flow receives is not ensured.

Prior work addresses these challenges separately. One line of work repurposes
the USB Extensible Debug Capability (xDBC) as a host-to-host
link~\cite{west2023real, njavro2025real}. It unifies networking and device
I/O on one bus, but it is limited to bulk transfers and offers no
differentiated service. A second line of work provides interrupt-based
differentiated service for USB device I/O~\cite{usb-diff-services}. It serves
each flow through its own completion path, but it targets bulk transfers to
USB~2 devices, not host-to-host communication. No existing system offers
differentiated host-to-host communication over USB~3.x with both isochronous
and bulk transfers.

This paper closes that gap. We present a real-time xDCI gadget driver for the
Quest RTOS~\cite{vcpu} that lets one machine appear to another as a USB
device with a configurable number and type of endpoints. Isochronous
endpoints carry periodic flows on a reserved schedule. Bulk endpoints carry
reliable data at high throughput. On the host side, USB interrupters route
each flow's completions to its own I/O VCPU, a bandwidth-preserving 
% server that isolates 
server, isolating
the flow from everything else on the bus. 
% To our knowledge,
Ours is the first system to use multiple USB interrupters to support
host-to-host communication with both isochronous and bulk transfers at
USB~3 rates. Bus latencies are predictable down to the 125~$\mu$s microframe,
which is sufficient for many CPS control operations.

This paper makes the following contributions:
\begin{itemize}
  \item A real-time xDCI gadget driver that provides host-to-host
    communication over USB~3.x with a configurable number and type of
    endpoints, including the frame alignment that isochronous service
    requires.
  \item The integration of interrupter-based flow differentiation with
    Quest's I/O VCPU scheduling, extending differentiated service from
    device I/O~\cite{usb-diff-services} to communication between hosts.
  % \item An evaluation showing that isochronous and bulk flows share one bus
  %   while each maintains its required service.
  \item An evaluation showing that isochronous and bulk flows share a bus
    while maintaining required service.
\end{itemize}

The rest of the paper is organized as follows. Section~\ref{sec:background}
motivates the use of USB in the context of a next-generation automotive E/E
architecture. This is followed by a primer on USB, the Quest RTOS, and differentiated
service.
Section~\ref{sec:design} presents the design and implementation of the
gadget driver. Section~\ref{sec:eval} evaluates the system, while
Section~\ref{sec:related} surveys related work, and
Section~\ref{sec:conclusion} draws conclusions.
   % 1
\section{Background}
\label{sec:background}

% Primer ONLY -- just what the reader needs to follow the Design (sec:design).
% Comparative positioning vs. prior art lives in Related Work (sec:related).
% Three primers, in dependency order: USB -> Quest RTOS -> Differentiated Service.
% \subsection{Case Study: An SDV E/E Architecture}
% \label{sec:bg:sdv}

Centralized Software-Defined Vehicle (SDV) architectures consolidate
functions that were once spread across many fixed-function electronic
control units onto a small number of hosts. Consolidation reduces
wiring, cost, and weight, and lets vehicle behavior be defined and
updated in software. It also makes each host responsible for two
distinct kinds of traffic at once: (1) inter-host communication, and
(2) I/O with the vehicle's devices. As a host now concentrates
safety-critical functions, the architecture pairs a primary machine with
a backup, keeping the two synchronized over a network,
so the backup takes over on failure. The same host must also handle input data from sensors, including cameras, LiDARs, and radars, execute control
functions, and subsequently deliver output signals to actuators, including motors, switches, and servos. Both inter-host and host-device traffic classes are subject to latency contraints, that must be met for safe vehicle operation.

\begin{figure}[!htb]
  \centering
  \includegraphics[width=\columnwidth]{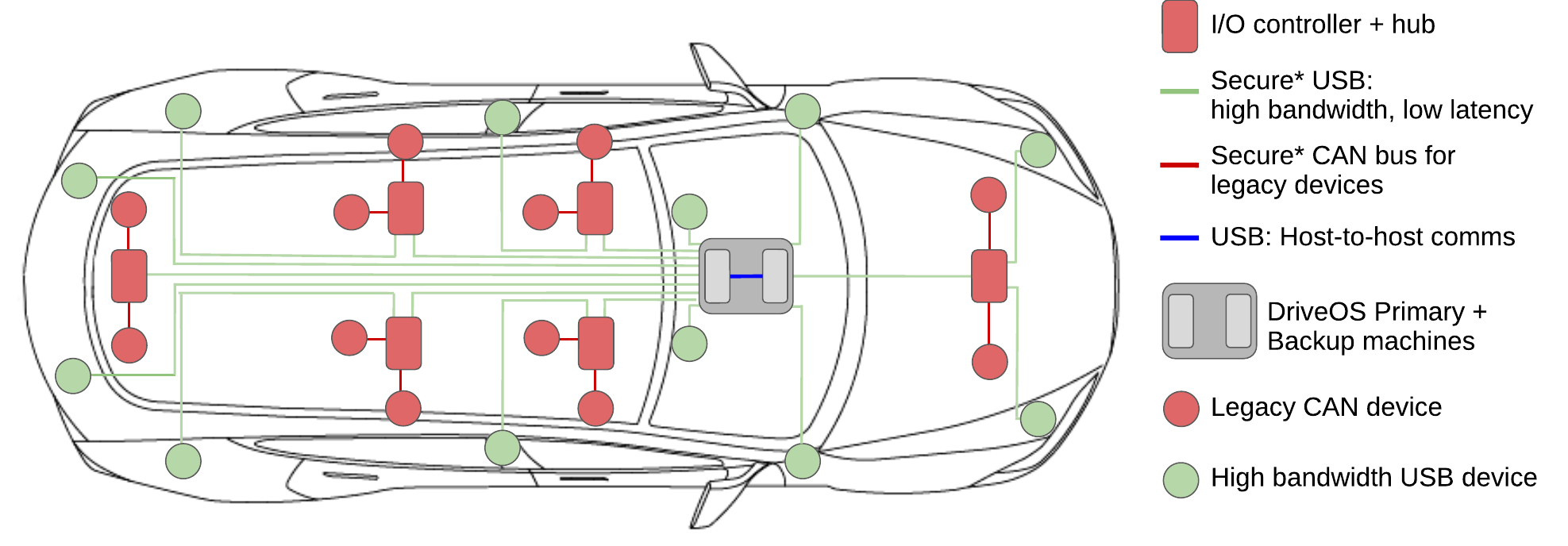}
  \caption{Example use-case: The Drako DriveOS$^{TM}$ vehicle management system E/E architecture.}
  \Description{Example use-case: The DriveOS software-defined vehicle management system.}
  \label{fig:driveos}
\end{figure}

Figure~\ref{fig:driveos} shows one such system, the DriveOS~\cite{driveos}
electrical/electronic (E/E) vehicle management architecture used by Drako Tech~\cite{drako_tech}. Here, a
single bus type carries both the host-to-host network and the device I/O data. Specifically, the design
uses USB for all major bus communication. This common bus solution requires different traffic types to be carefully coordinated, to ensure per-flow service guarantees in terms of throughput and delay. The rest of this section builds up the
pieces required to achieve this, beginning with USB itself.

\subsection{Universal Serial Bus}
\label{sec:bg:usb}

The Universal Serial Bus connects peripherals to a host. The host initiates every transfer, and a device sends data only when the host makes a request. This asymmetry is the defining property of USB. It keeps the device simple
and places the work of scheduling traffic on the host. A device presents its
data through \textit{endpoints}. An endpoint is a logical stream that moves data
in one direction, either to or from the host, and it is the unit at which the
bus addresses and reserves bandwidth. Successive versions of the standard have
raised the bus's signaling rate, from 12~Mbps in the first release, to 480~Mbps in USB 2.0, 5~Gbps in USB 3.0, 20~Gbps in USB 3.2, and 40~Gbps in USB~4. Its growth is what makes USB the bus of choice
for high-rate CPS sensors such as cameras and LiDARs.

Every transfer on the bus belongs to one of four types: (1) \textit{Control}
transfers configure a device, (2) \textit{Bulk} transfers carry large data reliably,
with no timing guarantees, (3) \textit{Isochronous} transfers handle rate-based data
on a reserved schedule and tolerate loss, and (4) \textit{Interrupt} transfers are for
small, latency-bounded messages. Each endpoint is associated with exactly one transfer type. For example, a
keyboard uses interrupt, a mass-storage device uses bulk, and a camera uses
isochronous. The rest of this paper concerns the two types that carry CPS
payloads, notably isochronous and bulk. Control appears only to configure a device, and
interrupt transfers play no further part.

Isochronous transfers are the natural carrier for periodic CPS flows. USB
divides time on the bus into 125\,$\mu$s \textit{microframes}. The host admits
an isochronous endpoint only when it successfully reserves the endpoint's requested
bandwidth, for a specific service interval, which is a fixed multiple of
the microframe. An
isochronous transaction carries no handshake and is never retransmitted. The
controller keeps to this schedule and drops what fails, so an error is isolated
and the next service instance carries the freshest sample of data. For a real-time stream
this makes sense when timely and fresh data is worth more than late information~\cite{mk-firm,west2004dwcs}. Any reliability required by an isochronous flow is added in
software.

The protocol reserves the interval, but it does not on its own deliver a usable
stream. As Section~\ref{sec:design:device} shows, an endpoint that starts on the
wrong frame is offered its bandwidth and still sends nothing. Realizing an endpoint's service 
guarantee is the work of the device driver.

Before any of this traffic can flow, the host must learn what the device offers.
Each device describes itself through a hierarchy of \textit{descriptors} that
the host reads during enumeration. They run from the device, through its
configurations and interfaces, down to its endpoints. The endpoint descriptor is
the one that specifies a flow's service in terms of transfer type,
direction,  maximum packet size, and service interval. An
isochronous endpoint uses the last two fields to establish a flow's transfer rate and latency bound, down to the number of bytes that may be sent in each interval. Our design
treats these parameters as a flow's service class.

\begin{figure}[t]
  \centering
  \includegraphics[width=0.5\columnwidth]{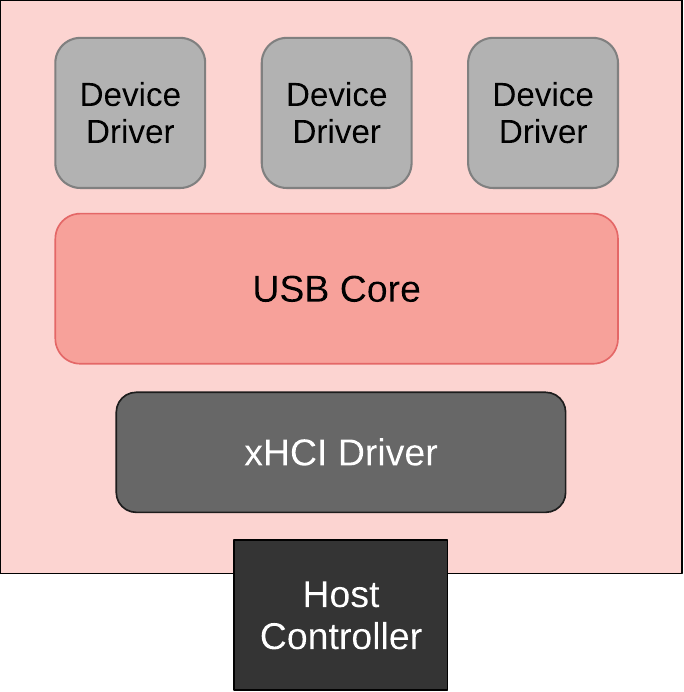}
  \caption{The USB host-side software stack: per-device class drivers access the host controller via the USB core and xHCI driver.}
  \Description{The USB host-side software stack: per-device class drivers access the host controller via the USB core and xHCI driver.}
  \label{fig:usb-stack}
\end{figure}

Most of the protocol logic lives on the host. 
Figure~\ref{fig:usb-stack} shows the host stack. 
A \textit{device driver} sits at the top and presents one device
to the rest of the system. The USB \textit{core} binds these per-device drivers
to the controller logic in the OS. The \textit{host controller driver} below it
operates the controller hardware. At the bottom is the physical host controller,
usually reached over PCIe. For USB~3 this controller follows the Extensible Host
Controller Interface (xHCI)~\cite{xhci-spec}, which defines how the controller
and the OS communicate.

Software and the host controller exchange work through a small set of
producer-consumer rings in system memory. There are three ring types:
(1) a single {\em command ring} for host commands, such as enabling a
device slot or configuring an endpoint, (2) a {\em transfer ring} per active endpoint, which holds the transfers queued for it, and (2) one or more {\em event rings} used by the controller to report completion and bus events, which may be raised with interrupts. The
unit of work on every ring is the Transfer Request Block (TRB). A
transfer descends through three stages before it reaches the wire. First, a
device driver submits a USB Request Block (URB) to the core. Second, the host
controller driver breaks the URB into one or more Transfer Descriptors
(TDs). Third, each TD is enqueued on the endpoint's transfer ring as a chain
of TRBs for the controller to execute. The device side reuses this
same machinery, and our design builds directly on it, as shown later in Section~\ref{sec:design:xdci}.

Commodity operating systems such as Linux and Windows build their USB stacks for
breadth and ease of use. They support as many devices as possible, and in doing
so they leave the real-time potential of the standard untapped. The result is
jittery, unpredictable delivery~\cite{west2023real, usb-diff-services}. With careful co-design between the USB
stack and the OS, the bus is able to support predictable I/O. That co-design rests on a
real-time foundation, which we describe next.

% Cover only what later sections lean on:
%   - Host/device roles, bus topology, endpoints, the transfer model.
%   - Transfer types: control, bulk, interrupt, isochronous -- and which map to
%     CPS traffic (ISOC for periodic/control, BULK for throughput).
%   - Frame/microframe schedule: the 125 us microframe granularity.
%   - USB 3 SuperSpeed: line rate, dual-simplex (why throughput is high).
%   - Host vs. device controller interfaces: xHCI (host) vs. xDCI (device). xDCI
%     full device emulation is the differentiator -- give it room here. [cite:
%     USB 3.2 spec; xHCI spec; xDCI] -- author adds keys to references.bib.

% Single-column figure (regular `figure`, \columnwidth) -- stays within one
% column.

\subsection{The Quest Real-Time Operating System}
\label{sec:bg:quest}

\begin{figure}[hbp!]
  \centering
  \includegraphics[width=\columnwidth]{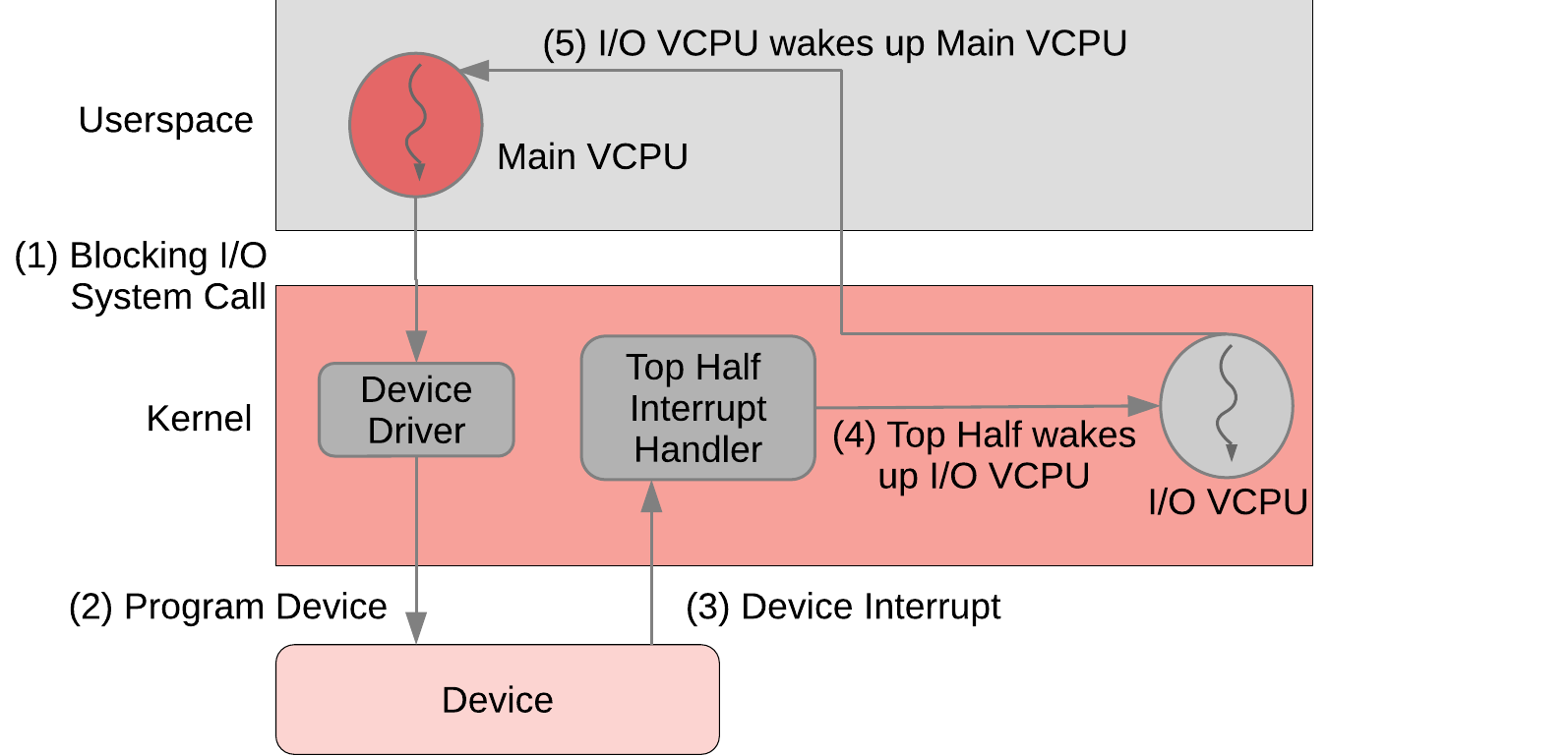}
  \caption{Interrupt handling in the Quest RTOS.}
  \label{fig:quest-int-handling}
  \Description{Interrupt handling in the Quest RTOS.}
\end{figure}

% Just enough of Quest to follow the design and the model (sec:modeling):
%   - The VCPU abstraction; Main VCPUs vs. IO VCPUs.
%   - How tasks/threads and interrupt (bottom-half) work map onto VCPUs.
%   - The scheduling/temporal-isolation guarantee that makes I/O predictable.
%     Forward-ref the gadget driver's place in this framework (sec:design).
%     NOTE: Quest specifics are authoritative from the authors -- keep claims
%     exact. [cite: Quest papers] -- author adds keys to references.bib.

Quest is a real-time operating system that schedules tasks on sporadic servers~\cite{sporadic-server}, and threaded interrupt handlers on simpler priority-aware bandwidth preserving servers, both implemented as Virtual-CPUs (VCPUs)~\cite{vcpu}. The two kinds of VCPUs in Quest are: (1) Main VCPUs that have a budget, $C$, every period,
$T$, and run application and system tasks, and (2) I/O VCPUs that handle device
interrupts. The I/O VCPUs simply have a utilization requirement, and are prioritized according to the task issuing the I/O request that led to the interrupt requiring service. 

In Quest, as in Linux, interrupt handling is split into 
% a top half and a bottom half. 
top and bottom halves. 
The top half acknowledges the interrupt and defers the rest of the work to a bottom half.
The two systems differ in where the bottom half then runs. Linux runs it as a softirq or
kernel thread with no bandwidth budget, so heavy interrupt load can starve other
work. Quest instead schedules a bottom half thread on an I/O VCPU, which is a bandwidth-preserving server~\cite{sprunt1989aperiodic,mercer1994reserves}
with utilization factor $U_{IO}$. When scheduled, the I/O VCPU is dynamically
assigned the period of the Main VCPU of the task awaiting I/O. 

% Include if space is available:
% Significantly, an I/O VCPU has a single replenishment, which is available at
% a future eligibility time. This guarantees its bandwidth utilization never
% exceeds UIO on the underlying physical CPU (or core). This approach is
% effective for shortlived interrupt handlers, as it avoids frequent
% reprogramming of replenishment timers needed to accurately manage budget
% usage of a Sporadic Server~\cite{mixed-crit-io}.

% taken from USB Diff services paper
Figure~\ref{fig:quest-int-handling} shows the mechanism. A task on a Main VCPU
makes a blocking I/O call, and the kernel programs the device. When the device
interrupts, the top half wakes the matching I/O VCPU and sets its period
$T_{IO}$ to the smallest period $T_{Main}$ among all Main VCPUs waiting on I/O.
Quest 
% schedules every VCPU by 
uses 
Rate-Monotonic Scheduling~\cite{rms}, so a
smaller period means a higher priority. It then sets the I/O VCPU's budget to
$U_{IO} \times T_{Main}~|~U_{IO}{\leq}1.0$, preserving bandwidth $U_{IO}$ over the Main VCPU's
period. When the I/O VCPU finishes the bottom-half thread, it blocks and wakes
the waiting Main VCPU. 

% The allocation of multiple I/O VCPUs for different interrupters to be added in
% the next subsection. 

\subsection{Differentiated Service}
\label{sec:bg:diffserv}

% Frame the problem this paper solves:
%   - The DiffServ idea (networking): traffic classes get per-class treatment.
%   - The prior interrupt-based differentiated service for I/O devices that this
%     work extends (the abstract's "earlier work").
%   - The problem statement: independent flows sharing one USB bus need per-flow
%     / per-class service guarantees -- set up the gap the Design fills. [cite:
%     DiffServ; prior interrupt-differentiation work] -- author adds keys.

\begin{figure}[htbp!]
  \centering
  \includegraphics[width=\columnwidth]{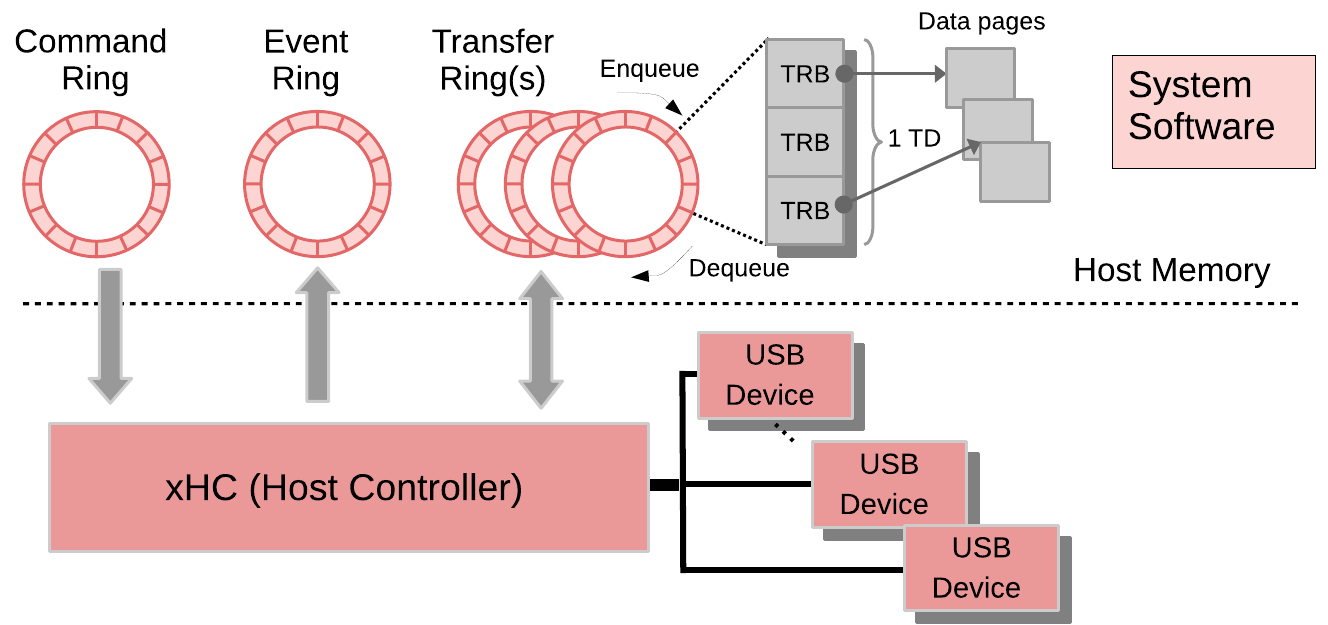}
  \caption{USB Interrupt Handling in Linux: a single interrupter has one event ring for all transfer completions. High priority traffic is potentially delayed by lower priority traffic whose interrupt on completion events are earlier in the ring. }
  \label{fig:linux-no-diffserv}
\end{figure}

A single USB bus rarely carries only one kind of traffic. On a consolidated CPS
host, a camera stream, a bulk file transfer, and a control loop may share one
controller. Each requires a different quality of service. The camera needs
sustained bandwidth, the control loop needs a tight bound on delay, while the file
transfer has no timing requirement but requires lossless communication. There is no guarantee of end-to-end service, combining both bus and host movement of data, when multiple flows share a single transfer completion
path. In such cases, a completed transfer raises
an interrupt and one handler services the completions of every flow. As a result, 
a low-priority transfer might be handled ahead of a high-priority one, leading to delay of an important traffic flow.
This priority inversion persists even when the handler inherits the highest
priority among the tasks awaiting I/O, since one handler cannot serve two
priorities at once~\cite{usb-diff-services}. Differentiated service is the
principle that each flow should instead be served according to its own
requirement, isolated from whatever else shares the bus.

Figure~\ref{fig:diffserv-timeline} illustrates the need for multiple
interrupters on a schedule of three tasks, $\tau_H > \tau_M > \tau_L$ in
priority. With a single interrupter, the low-priority flow's bottom half is
serviced ahead of the high-priority one, and the resulting inversion makes
$\tau_M$ miss its deadline. Giving each priority its own interrupter removes
the inversion and every task meets its deadline.

% --- Differentiated-service timeline (Fig. fig:diffserv-timeline) ----------
% Self-contained TikZ helpers (prefixed \ti...) for the two timing panels.
% \tiLane{y}{label}        one task/handler lane: left label + baseline.
% \tiTS{x1}{x2}{y}         a time-slice block on lane baseline y.
% \tiTSL{x1}{x2}{y}{txt}   ... with a centred label (used to mark L/H work).
% \tiReady{x}{y}           up arrow: task becomes ready to run.
% \tiIO{x}{y}              red down arrow: blocking I/O request.
% \tiDdl{x}{y}             red dashed vertical: deadline.
\newcommand{\tiLane}[2]{%
  \node[anchor=east,font=\scriptsize] at (-0.2,{#1+0.275}) {#2};
  \draw[black,thin] (0,#1) -- (\tiTmax,#1);
  \foreach \tx in {0,...,11} \draw[black,thin] (\tx,{#1-0.1}) -- (\tx,{#1+0.1});}
\newcommand{\tiTS}[3]{%
  \fill[blue!40] (#1,#3) rectangle (#2,{#3+0.55});
  \draw[black,thin] (#1,#3) rectangle (#2,{#3+0.55});}
\newcommand{\tiTSL}[4]{\tiTS{#1}{#2}{#3}%
  \node[font=\tiny] at ({(#1+#2)/2},{#3+0.275}) {#4};}
\newcommand{\tiReady}[2]{%
  \draw[-{Stealth[length=3pt,width=3pt]},semithick] (#1,#2) -- (#1,{#2+0.85});}
\newcommand{\tiIO}[2]{%
  \draw[-{Stealth[length=3.5pt,width=4pt]},red,very thick] (#1,{#2+0.9}) -- (#1,{#2+0.05});}
\newcommand{\tiDdl}[2]{%
  \draw[red,densely dashed,semithick] (#1,#2) -- (#1,{#2+0.9});}
\def\tiTmax{11}

\begin{figure*}[t]
  \centering
  % ---- shared legend (above the panels) -----------------------------------
  \begin{tikzpicture}
    \fill[blue!40] (0,0) rectangle (0.35,0.22);
    \draw[blue!65,thin] (0,0) rectangle (0.35,0.22);
    \node[anchor=west,font=\scriptsize] at (0.45,0.11) {Time Slice};
    \draw[-{Stealth[length=3pt]},semithick] (2.55,-0.02) -- (2.55,0.3);
    \node[anchor=west,font=\scriptsize] at (2.7,0.13) {Ready};
    \draw[-{Stealth[length=3.5pt]},red,very thick] (4.0,0.3) -- (4.0,-0.02);
    \node[anchor=west,font=\scriptsize] at (4.15,0.13) {I/O Request};
    \draw[red,densely dashed,semithick] (6.2,-0.02) -- (6.2,0.3);
    \node[anchor=west,font=\scriptsize] at (6.35,0.13) {Deadline};
  \end{tikzpicture}

  \par\medskip
  % ---- (a) single interrupter: priority inversion -> deadline miss --------
  \begin{subfigure}[t]{0.49\textwidth}
    \centering
    \begin{tikzpicture}[x=0.5cm,y=0.55cm]
      % tau_L: ready, run, I/O request, far deadline (met)
      \tiReady{0}{3} \tiTS{0}{1}{3} \tiIO{1}{3} \tiDdl{11}{3}
      % tau_H: ready, run, I/O request; resumes only after both BHs (delayed)
      \tiReady{2}{5} \tiTS{2}{3}{5} \tiIO{3}{5} \tiTS{7}{8}{5} \tiDdl{8}{5}
      % tau_M: ready at 2 but blocked until 8; overruns its deadline at 9
      \tiReady{2}{4} \tiTS{8}{10}{4} \tiDdl{9}{4}
      \node[red,font=\scriptsize\bfseries] at (9.45,4.78) {$\times$};
      % top half acks both completions
      \tiReady{4}{2} \tiTS{4}{5}{2}
      % single bottom-half lane: low-priority work (L) served before high (H)
      \tiTSL{5}{6}{1}{L} \tiTSL{6}{7}{1}{H}
      % lanes drawn last so the baseline shows over the time-slice boxes
      % lanes: tauH=5 tauM=4 tauL=3 TopHalf=2 BottomHalf=1
      \tiLane{5}{$\tau_H$} \tiLane{4}{$\tau_M$} \tiLane{3}{$\tau_L$}
      \tiLane{2}{Top Half} \tiLane{1}{Bottom Half}
      % numbers on the bottom-most lane only; x-axis label
      \foreach \x in {0,5,10} \node[anchor=north,font=\tiny] at (\x,0.85) {\x};
      \node[anchor=north,font=\scriptsize] at (5.5,0.35) {Time};
    \end{tikzpicture}
    \label{fig:diffserv-single}
  \end{subfigure}
  \hfill
  % ---- (b) one interrupter per priority: no inversion, no miss ------------
  \begin{subfigure}[t]{0.49\textwidth}
    \centering
    \begin{tikzpicture}[x=0.5cm,y=0.55cm]
      % tau_L: same as (a)
      \tiReady{0}{3} \tiTS{0}{1}{3} \tiIO{1}{3} \tiDdl{11}{3}
      % tau_H: resumes one slot earlier, deadline met
      \tiReady{2}{5} \tiTS{2}{3}{5} \tiIO{3}{5} \tiTS{6}{7}{5} \tiDdl{8}{5}
      % tau_M: runs 7..9, meets deadline at 9
      \tiReady{2}{4} \tiTS{7}{9}{4} \tiDdl{9}{4}
      % top half acks both completions
      \tiReady{4}{2} \tiTS{4}{5}{2}
      % high-priority bottom half runs first; low-priority one deferred to slack
      \tiTSL{5}{6}{1}{H} \tiTSL{9}{10}{0}{L}
      % lanes drawn last so the baseline shows over the time-slice boxes
      % lanes: tauH=5 tauM=4 tauL=3 TopHalf=2 BH(H)=1 BH(L)=0
      \tiLane{5}{$\tau_H$} \tiLane{4}{$\tau_M$} \tiLane{3}{$\tau_L$}
      \tiLane{2}{Top Half} \tiLane{1}{Bottom Half (H)} \tiLane{0}{Bottom Half (L)}
      % numbers on the bottom-most lane only; x-axis label
      \foreach \x in {0,5,10} \node[anchor=north,font=\tiny] at (\x,-0.15) {\x};
      \node[anchor=north,font=\scriptsize] at (5.5,-0.65) {Time};
    \end{tikzpicture}
    \label{fig:diffserv-multi}
  \end{subfigure}

  \caption{Differentiated service eliminates the priority inversion of a single
  USB interrupter. Both panels run the same arrivals: $\tau_L$ then $\tau_H$
  each issue a blocking I/O request and $\tau_M$ becomes ready. (a) Single interrupter: $\tau_L$'s bottom half (L) is serviced before $\tau_H$'s (H). The inversion delays $\tau_H$ and $\tau_M$ misses its deadline ($\times$). (b) One interrupter per priority: $\tau_H$'s bottom half (H) runs first and $\tau_L$'s (L) is deferred to slack. No inversion; every task meets its deadline.}
  \label{fig:diffserv-timeline}
  \Description{Figure showing how a single interrupter causes priority inversion,
  which is eliminated with multiple interrupters.}
\end{figure*}

\begin{figure}[htbp!]
  \centering
  \includegraphics[width=\columnwidth]{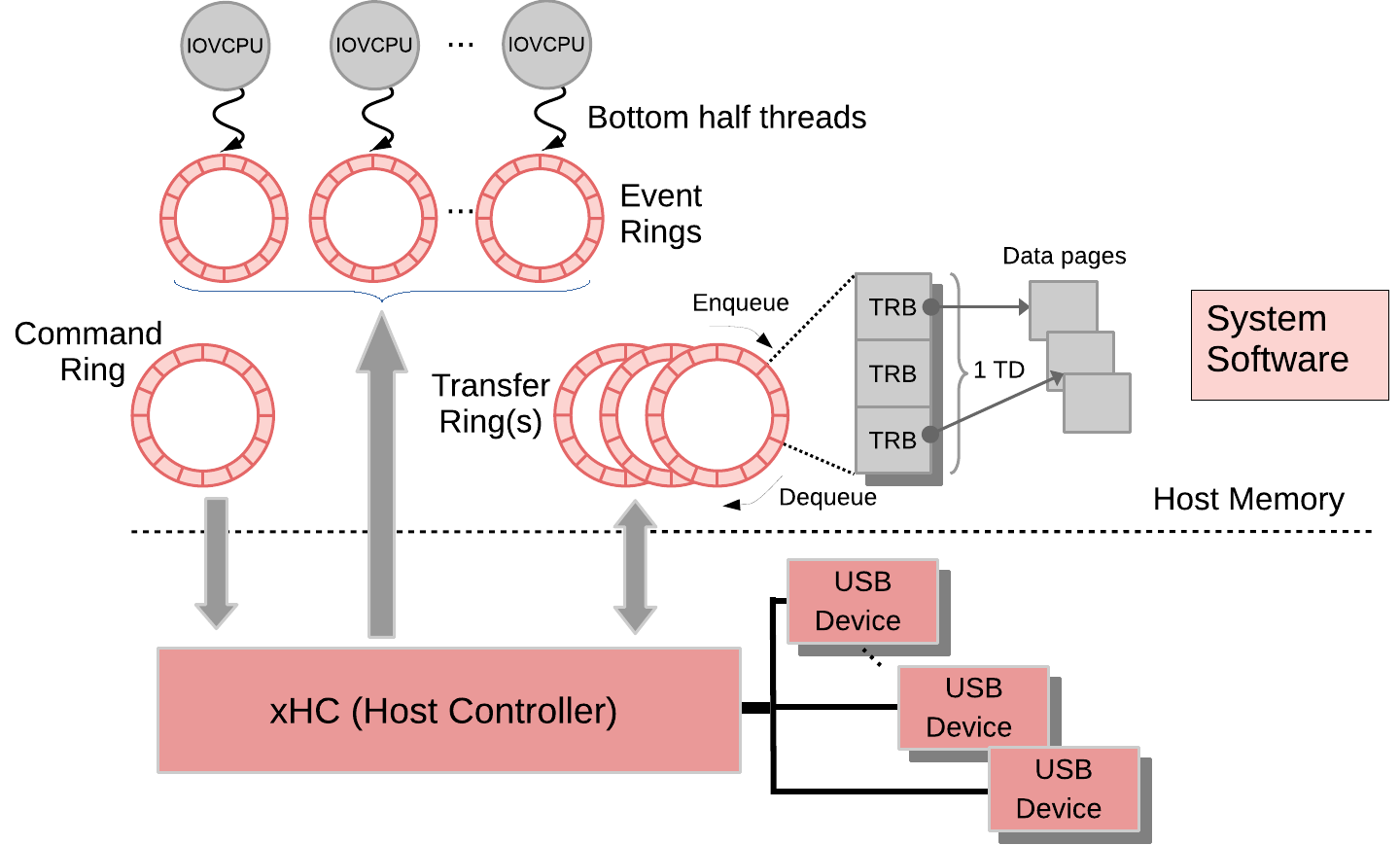}
  \caption{USB Interrupt Handling in Quest: Multiple interrupters, coupled with 
  priority-aware bottom-half scheduling allows for differentiated service. Each interrupter is associated with a separate event ring, avoiding 
  priority inversion.
  % the priority inversion problems of a single event ring.
  }
  \label{fig:quest-diffserv}
\end{figure}

% Interrupter / MSI mechanism, folded in below (moved here from the USB primer).
Earlier work removes the coupling by giving each class its own completion
path~\cite{usb-diff-services}. USB host controllers expose
\textit{interrupters}, a hardware facility for routing transfer completion
events to separate handlers. Interrupters are built on the controller's Message
Signaled Interrupt (MSI) capability. A single controller supports as many as
1024 of them, each with its own event ring and bottom-half
handler~\cite{xhci-spec}. Each interrupter is paired with its own Quest I/O VCPU
(Section~\ref{sec:bg:quest}), a priority-inheritance, bandwidth-preserving
server, as shown in Figure~\ref{fig:quest-diffserv}. This pairing charges a flow's completions to that server alone and
isolates it from the rest. The utilization factor $U_{IO}$ is the bandwidth the
server reserves for the flow. A larger $U_{IO}$ gives the flow a larger budget
each period, and so more throughput. For bulk transfers, a task tunes $U_{IO}$
to the service it needs.

Isochronous traffic cannot be tuned in this way. Its service level is a fixed
property of the flow, not a free parameter. A camera fixes its frame rate, and a
control loop fixes its period. USB encodes these as a service interval and a
number of bytes per interval, set when the endpoint is admitted and unchanged
thereafter. A task cannot choose an operating point here. The system must meet
the requirement it is given. The framework still works the same way, with each
flow on its own server. Its goal changes from tuning throughput and delay to
guaranteeing the service level the flow requires. The rest of this paper shows
how to provide this guarantee for both isochronous and bulk traffic.
     % 2  primer only (related work is separate)
\section{Design and Implementation}
\label{sec:design}

% Merged Design + Implementation. Within EACH subsection, lead with the design
% rationale ("what / why"), then the implementation specifics ("how, in Quest":
% data structures, kernel touch-points, LOC). No separate implementation block.
%
% Reading order (dependency): foundation (xDCI) -> the two driver halves
% (device, host) -> the scheduling glue (I/O-VCPU) -> end-to-end synthesis.
%
% ORIENT FIRST: introduce the end-to-end data-path figure (fig:e2e) here, at the
% top of the section, and refer back to it as each component appears -- even
% though the detailed flow walkthrough is the last subsection (sec:design:flow).
% Keep the experimental hardware/platform in the Evaluation setup, not here.
%
% Wide landscape diagram -> spans both columns (figure*). Scaled to 0.88\textwidth.
% Negative \vspace pulls it up into the top margin so it sits higher on the page.
\begin{figure*}[t]
  \vspace*{-22pt}
  \centering
  \includegraphics[width=0.88\textwidth]{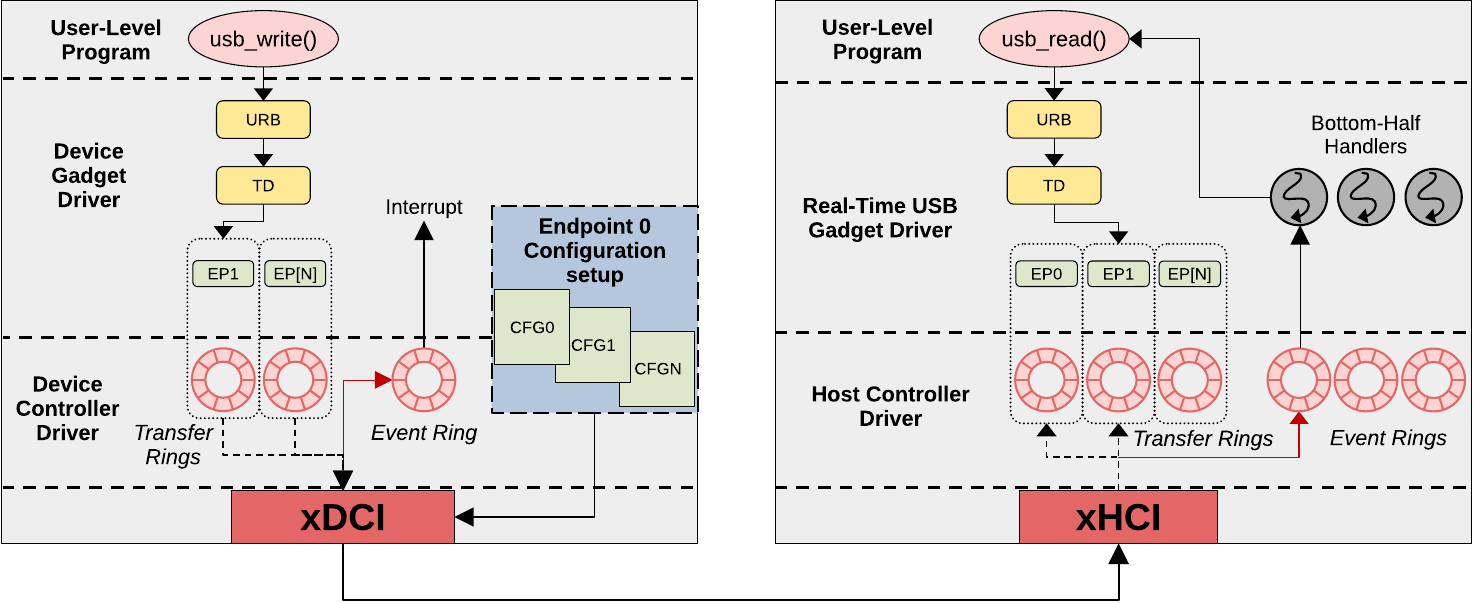}
  \caption{System perspective: the end-to-end data path from a user program on
    the sending host, through the device-side xDCI gadget driver, across the
    USB~3 bus, to the host-side driver and the receiving user program.}
  \label{fig:e2e}
  \Description{System perspective: the end-to-end data path from a user program
  on the sending host, through the device-side xDCI gadget driver, across the
  USB~3 bus, to the host-side driver and the receiving user program.}
\end{figure*}

\subsection{xDCI Architecture}
\label{sec:design:xdci}

Extensible Device Controller Interface (xDCI) is an interface specification for the USB device controllers (UDC) that exposes the programming-model for the UDC to the host OS. While traditionally UDCs were often embedded within the I/O devices such as keyboards, webcams, and flash drives, it was the smartphone market that introduced xDCI controllers more broadly through dual role devices (DRDs) that could act both as a USB host and a device. Such DRD configuration was quickly adopted in the broader SoC market with most of the new Intel SoCs coming equipped with a dedicated xDCI controller. However, having an xDCI controller itself is not a sufficient requisite since the OEM has to explicitly connect and configure the USB-C ports and subsystem to support device mode configuration. Our work is based on Synopsys DesignWare USB3 (DWC3) SuperSpeed Device Controller \cite{synopsys-dwc3-pg}. In this paper, we specifically focus on Intel's 13th generation Raptor Lake processors that feature DWC3 support. However, certain processors as early as the Intel Atom ``Bay trail'' family from 2013, right up to forthcoming ``Nova Lake'' processors may include DWC3 capabilities. The same controller is also found in multiple ARM processors too.

At the architectural level, the DWC3 device controller exposes its functionality through the same family of in-memory data structures introduced for xHCI in the previous section, namely transfer rings built from TRBs, a command ring, and an event ring. This symmetry is deliberate since the device side reuses the host-side transfer abstractions so that buffers are still described by TRBs linked into rings. Ownership is still arbitrated by a hardware-owned bit, and a logical transfer is still expressed as a Transfer Descriptor. We therefore focus on the aspects of the programming model that are particular to operating as a device rather than a host.

The first such aspect is endpoint management. The controller provides a preset
number of unidirectional physical endpoint resources, and software maps the USB
endpoints defined by the device's descriptors onto these physical endpoints
during enumeration. Software always operates in terms of physical endpoints,
with the controller translating between USB and physical endpoint numbers as
packets move in each direction. A USB control endpoint is a notable case, since
its bidirectional nature requires two physical endpoints, one for each
direction. Configuration, transfer initiation, and teardown are driven by
endpoint configuration and start transfer commands, which form the primary control path
between the gadget (device) software and hardware.

The second aspect is how the device reports progress back to software. 
% Whereas 
While
a host schedules and initiates traffic, a device must respond to a host that owns the bus, so completions and bus events arrive asynchronously and are delivered through the event buffer, a ring in system memory into which the controller writes events before raising an interrupt. Events fall into two categories: endpoint-specific events, which report the progress and completion of transfers on individual endpoints, and device-specific events, which report global conditions such as USB reset, connection, and link state transitions. The latter category has no real counterpart on the host side and is what drives the device's enumeration and power-state handling.

Together, the endpoint command interface and the event buffer specialize the shared TRB-based transfer machinery for device operation, and the same machinery underlies every transfer type. Bulk and interrupt transfers map onto it directly, isochronous transfers add interval-aligned scheduling so that buffers are presented in the microframe the host expects, and control transfers layer the two and three-stage setup, data, and status sequence on top of it. 

A device's control endpoint is the only endpoint active at reset, and enumeration proceeds over it as a series of control transfers. The remaining endpoints do not exist until the host issues a SET\_CONFIGURATION request, at which point software responds with the endpoint-configuration commands that map physical endpoints onto the endpoints of the selected configuration. Bulk, interrupt, and isochronous traffic can flow only after this step. The control endpoint thus bootstraps the data plane using the same command and event interface that carries ordinary transfers.

% Elaborate beyond the sec:bg:usb primer -- the specific xDCI mechanisms the
% design exploits (this is the design-relevant view, do not re-do the primer):
%   - Device-controller model: TRBs, transfer rings, event rings, doorbells.
%   - Interrupters + event handling (the hook differentiated service hangs off).
%   - What "full device emulation" buys: all transfer types + USB 3 line rate.
%   - Hardware constraints that motivate the later design choices.
% Figure: xDCI register / ring / interrupter model.
% [verify: xDCI mechanism details]  [cite: xDCI / xHCI spec]

\subsection{Device Side xDCI Gadget Driver}
\label{sec:design:device}

% =============================================================================
% SCOPE (agreed): device-side EGRESS (IN endpoints). Target ~1.5 to 2 columns,
% concept level: keep the TRB / DEPCMD / HWO / ESIT / FIFO vocabulary, but drop
% function names and line numbers. Lead with design rationale, then mechanism.
% The egress datapath (P2) is the core of this subsection: make it unambiguous.
% Deferred elsewhere (do NOT cover here):
%   - I/O-VCPU admission / per-flow VCPU mapping  -> sec:design:iovcpu
%   - end-to-end flow + isoc cold-start ordering  -> sec:design:flow
%   - throughput numbers + cross-flow contention  -> Evaluation
% Optional figure: the egress datapath (arm-a-TRB -> doorbell -> DMA -> event).
% Consider \paragraph{} run-in headings if the seven units feel too flat.
% =============================================================================

% Our explanation of device-side xDCI gadget driver starts with 
We begin with a
brief overview of
USB SuperSpeed (SS) terminology and transfer dynamics. In USB SS, the host solicits 
IN data by sending an ACK Transaction Packet (ACK-TP) to the device, to which the device can
either reply with its Data Packets (DPs) or with a NRDY signal indicating it does not have
the desired data. Once the host receives NRDY, and unlike in the previous versions of USB, the
host stops polling the device thereby saving energy and it waits for the device to transmit 
ERDY signal on its own indicating that its endpoint has the data. Once ERDY is received, the host 
again sends ACK-TP and this time receives the data. For the case of isochronous endpoints, there 
are no ERDY/NRDY signals but rather the host explicitly requests the data once every \textit{bInterval}
and if it does not receive it on time, the packet is considered lost.

With this transfer model in place, we turn to the structure of the gadget driver,
which corresponds to the device half of Figure~\ref{fig:e2e}. We focus throughout
on \textit{egress}, the direction in which the device transmits to the host, which
is carried entirely on IN endpoints. The driver exposes several such endpoints, although in
our configuration (and without loss of generality) we define two isochronous and two bulk IN endpoints, alongside the
bidirectional EP0 control endpoint. Each is a separate physical endpoint,
identified by its Device Context Index (DCI), and is backed by its own ring of TRBs
in system memory onto which the driver stages outbound payloads. These endpoints do
not exist until enumeration, when the host configures them through EP0, a
provisioning step we return to at the end of the subsection.

One structural
property shapes everything that follows: the device controller reports the
completions of all its endpoints through a single shared event ring and
interrupter. Unlike the host side, which can steer different flows to separate
interrupters (Section~\ref{sec:design:host}), the xDCI on current SoCs cannot differentiate flows
by routing them to distinct interrupt vectors; it differentiates them entirely in
software, through the per-class service policy developed in the rest of this
subsection. However, there's no fundamental limitation preventing it from implementing
MSI-styled interrupters as xHCI does.

On the egress path it is the controller, not the CPU, that places bytes on the
wire, so the driver's role is to prepare work in memory and to signal the
controller that it is ready. It does so by writing a TRB into the target
endpoint's ring and issuing an endpoint command (DEPCMD) that directs the
controller to scan that ring. Preparing the TRB must follow a deliberate order: the
driver bounds the payload to the endpoint's per-interval budget (bytes per interval) and copies it into
a controller-visible DMA buffer, then records the buffer's address and length
together with the transfer's control bits in a free ring entry. This is followed by setting the
Hardware-Owned (HWO) bit only once the rest of the entry is in place, since it is
this bit that surrenders the slot to the controller. The command that follows
reflects the state of the transfer. A STARTTRANSFER command initiates a new transfer and returns a
resource index that the driver retains for reuse. An already-running transfer is
advanced with the lighter UPDATETRANSFER command that simply asks the controller to
re-scan the ring. Finally, an ENDTRANSFER command completes the transfer. The remainder of the
exchange is handled in hardware, so that at the host's next solicitation the controller
streams the armed buffer out of the endpoint's TX FIFO as a burst of Data Packets
and clears the slot's HWO bit to mark it complete. The device controller driver reclaims a
slot by observing the same bit it previously set.
%Because each ring is finite,
%a producer that arms entries faster than the controller retires them eventually
%finds the ring full. This situation is resolved in opposite ways by the two transfe%r classes.

The driver learns of completion through the device's single event ring: the
controller posts an event and raises its lone interrupter, and the driver services
it in a top-half/bottom-half split handler, whose bottom half advances each ring's consumed
position, freeing the entries that bulk transfers wait on and delivering the not-ready signal
that starts the next isochronous transfer. How that bottom half is scheduled is
deferred to Section~\ref{sec:design:iovcpu}.

The two classes share this datapath and diverge only in how they respond to a full
ring, and it is here, rather than in any dedicated mechanism, that the driver
expresses each flow's service class. An isochronous endpoint is rate-based and
tolerant of loss, so there is nothing to gain by blocking its producer: a payload
that misses its interval is worthless, and the link will not retransmit it. The
driver therefore lets isochronous writes shed data by dropping the payload and returning
when no ring entry is free, so that the producer moves on to a fresher sample rather
than stalling. To keep the controller supplied in spite of this, it carries a small
buffered sequence of transfers in the ring; in steady state the producer hands over a batch of
distinct payloads that the driver arms under a single doorbell, maintaining a non-empty buffer without re-sending the same sample. Bulk transfers are handled differently.
Because their delivery must be lossless, the driver cannot drop data, so a write to a full
ring instead waits, yielding the CPU and retrying until an entry is freed. The
transfer is kept open so that the ring streams continuously rather than restarting
with each payload. Back pressure is thus the point at which the two classes diverge:
rate-based flow guards its timeliness by discarding miss-timed transfers, while reliable flow guards its data by waiting until a free buffer slot is available for transmission.

% Single-column timing figure for the isoc frame-alignment finding (P6).
\begin{figure}[!htb]
\centering
\resizebox{\columnwidth}{!}{%
\begin{tikzpicture}[x=0.72cm,y=0.85cm,>=stealth,font=\small, % <<< FONT SIZE: change \small here to adjust all labels >>>
    poll/.style={->,thick},                                            % host poll/read (solid, above box)
    tx/.style={->,thick,dash pattern=on 2pt off 1.5pt,blue!25!black},  % device send (dashed, below box)
    cell/.style={draw,thick},
    hatch/.style={pattern=north east lines}]
  % --- legend (top, centered over x=0..12) ---
  \draw[poll] (1.5,5.1) -- (1.5,5.6);
  \node[anchor=west] at (1.75,5.35) {Host Polling};
  \draw[tx]   (7.0,5.1) -- (7.0,5.6);
  \node[anchor=west] at (7.25,5.35) {Device Transmission};
  % --- aligned (top) ---
  \begin{scope}[yshift=2.3cm]
    \node[anchor=east,font=\small\bfseries] at (-0.2,0.9) {aligned};
    \foreach \i in {0,4,8}
      \draw[cell,fill=dpgreen] (\i,0.55) rectangle (\i+2,1.25);
    \foreach \i in {2,6,10} \draw[cell] (\i,0.55) rectangle (\i+2,1.25);
    \foreach \cx in {1,5,9} {
      \draw[tx]   (\cx,0)    -- (\cx,0.55);
      \draw[poll] (\cx,1.25) -- (\cx,1.8);
      \node[font=\small\bfseries] at (\cx,2.05) {DP};
    }
    % --- microframe boundary ticks + numbers (0..6) at the box borders ---
    \foreach \mf in {0,1,2,3,4,5,6} {
      \draw[thick] (2*\mf,0.6) -- (2*\mf,0.38);   % small tick above the number
      \node[font=\small] at (2*\mf,0.18) {\mf};
    }
  \end{scope}
  % --- unaligned (bottom) ---
  \node[anchor=east,font=\small\bfseries] at (-0.2,0.9) {unaligned};
  \foreach \i in {0,4,8}  \draw[cell,fill=zbred] (\i,0.55) rectangle (\i+2,1.25);
  \foreach \i in {2,6,10} \draw[cell] (\i,0.55) rectangle (\i+2,1.25);
  \foreach \cx in {1,5,9} {
    \draw[poll] (\cx,1.25) -- (\cx,1.8);
    \node[font=\small\bfseries] at (\cx,2.05) {ZLP};
  }
  \foreach \cx in {3,7,11} \draw[tx] (\cx,0) -- (\cx,0.55);
  % --- microframe boundary ticks + numbers (0..6) at the box borders ---
  \foreach \mf in {0,1,2,3,4,5,6} {
    \draw[thick] (2*\mf,0.6) -- (2*\mf,0.38);   % small tick above the number
    \node[font=\small] at (2*\mf,0.18) {\mf};
  }
  \node at (6.0,-0.5) {microframe ($125\,\mu$s each)};
\end{tikzpicture}%
}
\caption{Isochronous Frame Alignment: Aligned (top): the Device Packets (DP)
land in the same microframe as the host's polls. Unaligned (bottom): Device
transmissions and host polling do not occur in the same microframe, leading to
the host receiving Zero-Length Packets (ZLP).}
\label{fig:isoc-align}
\Description{Isochronous Frame Alignment: Aligned (top): the Device Packets (DP)
land in the same microframe as the host's polls. Unaligned (bottom): Device
transmissions and host polling do not occur in the same microframe, leading to
the host receiving Zero-Length Packets (ZLP).}
\end{figure}
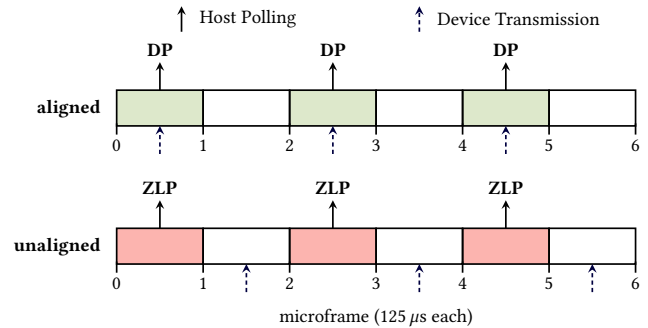

Isochronous transfers also differ from bulk in how they start. A bulk endpoint may
begin transmitting as soon as it has data, but an isochronous endpoint cannot start
on its own. It begins only when the controller reaches an interval it must service,
finds no data armed, and signals the driver to start. The driver responds with a
STARTTRANSFER, which must name the frame on which transmission begins. It reads the
current frame from the controller's running counter, adds a fixed margin so the
command is placed in time, and aligns the result down to the endpoint's service
interval. The alignment has no effect at the default 125\,$\mu$s interval, but it is
required at every slower rate. Figure~\ref{fig:isoc-align} contrasts the two cases. With an aligned start, the device's transmit opportunities depend on the host's polling rate
and each carries a data packet. With an unaligned start, every host poll returns a zero-length packet and the stream never recovers. Aligning
the target to the service interval eliminates
this and lets a second isochronous flow run at 250\,$\mu$s, 1\,ms, or 2\,ms. Two
failures are handled by retrying a later frame: a target that has already passed,
and a stale transfer resource that must first be torn down. A genuine miss is
reported as a per-endpoint event, which the driver uses as a direct deadline-miss
signal.

\subsection{Host Side xDCI Gadget Driver}
\label{sec:design:host}

% Host-side RX datapath (companion to device egress, sec:design:device). ~2 cols,
% concept level (URB/TRB/MSI/interrupter/double-buffer; no function names). Plan:
%   H1 two drivers + RX direction (host half of fig:e2e)
%   H2 down-path: arming the stream
%   H3 split interrupt + per-interrupter routing (THE differentiation; scheduling
%      deferred to sec:design:iovcpu)
%   H4 completion handler: self-sustaining resubmit loop (pool = slack) + deposit
%   H5 depth-2 latest-only double-buffer (lossy, latency-over-completeness; depth-3
%      = race-free) + busy-poll reader
% Skip: slab allocator, zero-buffer/magic, frame slack. Contention -> Evaluation.

On the host the same transfer machinery runs in the receive direction, and the
gadget's IN endpoints appear to a user program as files it can read. Two drivers
are stacked to make this work, both shown in the host half (right-hand side) of Figure~\ref{fig:e2e}.
The lower one is the xHCI host-controller driver, which owns the controller's
transfer and event rings, its doorbells, and its interrupts.
The host gadget driver sits above, which presents each of the device's
IN endpoints, the two isochronous and two bulk channels, as a file that a program
opens and reads. The gadget driver owns the per-channel buffers and the pool of in-flight
URBs that keep each channel supplied. Receiving data is therefore split into two halves:
a down-path, in which a read arms the host controller with requests for incoming data,
and an up-path, in which a completed transfer traverses from an interrupt back up to
the reader thread.

A read does not, in steady state, fetch a packet. The first time an isochronous channel is
opened, the gadget driver allocates its buffers and submits a pool of receive URBs at once; the host-controller driver turns each URB into a set of
transfer-ring TRBs and rings the endpoint's doorbell, so the controller holds many
incoming transfers queued before any data arrives. Allocation follows a \textit{lazy} approach: a channel
that no one has opened stays dark, with nothing submitted, so the controller never
streams into an endpoint that has no reader. Once the pool is armed, the read does
no submission of its own; it only inspects the per-channel buffer that the up-path
keeps filled. How the pool re-arms itself, so that the channel stays supplied
without a submission on every read, is the subject of the up-path below.

Completions return through the controller's interrupter. When an incoming transfer
finishes, the controller writes a transfer event to an event ring and raises an
MSI. The top-half handler 
% is deliberately small: it 
acknowledges the interrupter and
wakes the bottom-half that will service it, but does no ring work itself. The
bottom-half then drains the event ring, matches each event to its outstanding
transfer, and runs the completion handler for any transfer that has finished.

The host controller driver
assigns each flow to its own interrupter using the TRBs. This ensures the controller posts that flow's
events to a separate event ring, and each interrupter is drained by its own
bottom-half thread. Different flows, therefore, make progress in parallel, with each bottom-half
inheriting the priority from its originating VCPU user thread. On the device side, there is a single interrupter that funnels every
endpoint's events through one path, and flows are separated only by software policy
(Section~\ref{sec:design:device}).

The completion handler has two responsibilities for each finished transfer: it
returns the URB to the controller and it delivers the packet to the reader. The
former sustains the stream. Rather than wait for a subsequent read to submit fresh
work, the handler resubmits the completed URB immediately, requesting that the
controller schedule it in the next available frame, so that it rejoins the tail of
the in-flight pool. Submitted once at startup, the pool is thereafter
self-sustaining: it replenishes itself one URB at a time and remains full without
any submission on the read path.

The real-time behavior of this loop is a consequence of the pool's depth. Each completed URB must be
resubmitted before the remaining URBs are drained, otherwise the controller exhausts its armed
intervals. For example, by supporting a pool of 64 URBs, each with a 125~$\mu$s service interval, it would take 8\,ms before the buffer was completely drained. The bottom-half thread 
may therefore be scheduled many service intervals later, without the stream missing one; only
a starvation that outlasts the entire pool forfeits an interval. Per-packet
timeliness is thus decoupled from the latency of any individual wake-up, allowing a
real-time-scheduled host to keep a high-rate isochronous stream armed.

The buffer between the completion handler and the reader is not a queue but a
latest-only double-buffer. The handler, as the sole producer, writes each
received packet into the buffer's inactive slot, issues a memory fence, and then
publishes the packet by flipping an index to that slot and incrementing a
generation counter. The reader, as the sole consumer, busy-polls: when the
generation counter has advanced since its last read it copies from the published
slot, otherwise it returns immediately, so that it observes each deposit at the
finest granularity its schedule allows. A single writer and 
% a single 
reader are
coordinated only by the generation counter and the fence, rendering the path entirely
lock-free.

Because the buffer retains only the most recent packet, it is lossy by design.
A packet is overwritten the moment the handler publishes a newer one, and data is lost if the reader fails to complete the copying of a prior packet before the next one arrives. Loss therefore reflects the reader's lateness rather than any
error on the link, and a reader that polls faster than the deposit rate observes
every packet. This latest-only buffering of data trades completeness for
latency. It guarantees that the reader always obtains the freshest sample with no
queueing delay, and it keeps the receive path lock-free, at the cost of dropping
bursts that the reader cannot service in time. For a timing measurement, in which
the freshest sample is precisely the quantity of interest and a missed interval is
a clean, countable gap, this is the appropriate design; a workload requiring
lossless delivery would instead adopt a deeper FIFO and accept the resulting
latency and staleness. Two slots are the minimum that afford the reader a stable
region to read from, and a third would be required to eliminate a residual
torn-read window; the choice of two is therefore a deliberate preference for
latency over completeness.

\subsection{Integration with the Quest I/O-VCPU Scheduling Framework}
\label{sec:design:iovcpu}

Differentiated service is realized when each flow is given its own scheduling
reservation. Quest provides two reservation types, introduced in
Section~\ref{sec:bg:quest}: a Main VCPU, which grants an application thread a fixed
budget every period, and an I/O VCPU, which provides a service reservation for interrupt handling based on the task that generated the I/O request.

On
the host, a flow is carried by one of each reservation type. The reader thread runs on a Main VCPU,
and the bottom half that drains that flow's completions runs on a dedicated I/O
VCPU. The two are bound together along the path described in
Section~\ref{sec:design:host}: the reader tags its endpoint with an I/O VCPU class,
the class selects the interrupter to which the controller posts the flow's
completions, and that interrupter is drained by a bottom half bound to the matching
I/O VCPU. Each flow therefore has a completion path of its own, scheduled under a
budget of its own.

What turns these reservations into a service guarantee is the choice of their
budget and period, and both follow from the flow's latency requirement. That
requirement is the endpoint's service interval: a flow serviced every 125\,$\mu$s
must have its completions observed at that cadence, whereas one serviced every
2\,ms may be observed far less often. The period is chosen to oversample
its service interval, so that no deposit is overwritten before the reader copies
it, and its bottom half's I/O VCPU inherits the period of the most urgent reader on
its interrupter, by the rule of Section~\ref{sec:bg:quest}, so the kernel work runs
as urgently as the most demanding flow it serves. The budget, in turn, is sized to
the work each period performs: either a copy out of the double-buffer for the reader, or a
drain of the event ring for the bottom half. The period thus captures the flow's
latency requirement and the budget is set large enough to cover the processing cost. Smaller periods and larger budgets translate directly into larger CPU reservations.

These reservations are not granted unconditionally. Quest admits each against a
per-core utilization bound that tightens as the number of reservations on the core
grows, and a set that would exceed the bound is rejected rather than silently
overcommitted~\cite{vcpu}. Admission is what makes the budgets a guarantee: once a flow's
reader and bottom half are admitted, each receives its budget every period
regardless of the other flows on the core, and the bandwidth-preserving discipline
of the I/O VCPU keeps a busy interrupter from claiming more than its reserved share.
The bound also constrains how the parameters may be tuned. A flow's budget cannot be
cut without starving the copy it must perform, so additional flows are accommodated
by lengthening periods rather than by shrinking budgets, and a tighter latency
requirement is correspondingly a heavier admission cost.
%In the configuration
%evaluated in this paper, all reservations share a single core, which reserves each %flow its
%share of CPU time but not its interrupt-to-bottom-half wake latency;
%Section~\ref{sec:eval} examines what that distinction costs when flows contend.

The device side mirrors this arrangement only halfway. Producer threads are assigned to Main VCPUs, to ensure that transfer rates are maintained correctly, but interrupt processing is not separated across different I/O VCPUs as there is only one interrupter for the entire device controller.
         % 3  Design and Implementation (merged)
\section{Evaluation}
\label{sec:eval}

\subsection{Experimental Setup}
\label{sec:eval:setup}

We evaluate the system on a two-machine testbed. Each machine is a SuperMicro
X13SRN-H industrial board built around an Intel 13th-generation Raptor Lake SoC,
whose integrated DWC3 controller provides the SuperSpeed xDCI device interface that
our gadget driver targets. The two boards are identical: one acts as the device,
presenting itself through the xDCI gadget driver, and the other as the host, driving
the link through its xHCI host controller. Both run the Quest real-time operating
system. We pin every real-time flow on each machine to a single CPU core, which
removes cross-core interference and the latency it introduces. The two machines are
connected 
% directly 
by a SuperSpeed USB-C to USB-A cable.

% [verify setup: exact SoC SKU, RAM, Quest config/version, USB gen -- add if wanted]

\subsection{Throughput Measurements}
\label{sec:eval:exp1}

% TODO: narrate Figure~\ref{fig:exp1-timeseries} -- what the series shows (metric,
% flows, configuration) and the takeaway; add the Linux baseline where relevant.
% Frame as an RQ if helpful. Keep raw/processed data under results/.
% NOTE: rename this subsection to match what exp1 actually measures.

We begin by establishing the maximum sustainable link throughput, since this
figure bounds every workload the system carries. The relevant ceiling concerns isochronous endpoints, as they are granted a fixed budget of data in each multiple of the 
125\,$\mu$s microframe service interval. Consequently, the achievable throughput is set by how much
data that budget admits. We therefore measure achieved throughput as a function of
the endpoint's burst configuration, the parameter that determines the per-interval
budget.

\begin{table*}[!htb]
  \centering
  \footnotesize
  \caption{Isochronous throughput by burst configuration. For each scenario the
    table lists the configuration parameters, the theoretical maximum throughput,
    and the throughput measured over a 180\,s run (average, minimum, and maximum),
    together with the achieved fraction of the theoretical maximum. All throughputs
    are in Mbps.}
  \label{tab:throughput}
  \begin{tabular}{@{}rrrrrrrrrrr@{}}
    \toprule
    Scenario & bMaxBurst & Pkts/Burst & Mult & bInterval & Pkts/Interval &
      Theoretical & Measured Avg & Min & Max & Achieved (\%) \\
    \midrule
    1 & 0  & 1  & 0 & 1 & 1  & 65.536   & 65.0   & 65   & 65   & 99.2 \\
    2 & 1  & 2  & 0 & 1 & 2  & 131.072  & 131.0  & 131  & 131  & 99.9 \\
    3 & 3  & 4  & 0 & 1 & 4  & 262.144  & 260.3  & 259  & 261  & 99.3 \\
    4 & 7  & 8  & 0 & 1 & 8  & 524.288  & 518.4  & 517  & 520  & 98.9 \\
    5 & 15 & 16 & 0 & 1 & 16 & 1048.576 & 1032.1 & 1027 & 1035 & 98.4 \\
    \bottomrule
  \end{tabular}
\end{table*}

\begin{table*}[!htb]
  \centering
  \footnotesize
  \caption{Experimental VCPU parameters for different flow specifications.}
  \label{tab:vcpu_params}
  \begin{tabular}{@{}rrrrrr@{}}
    \toprule
    VCPU Type & Budget ($\mu$s) & Period ($\mu$s) & CPU Utilization & Endpoint Type & bInterval \\
    \midrule
    Main & 17 & 62 & 0.274 & Isochronous (ch0) & 1 (125~$\mu$s) \\
    Main & 17 & 1000 & 0.017 & Isochronous (ch1) & 16 (2000~$\mu$s) \\
    Main & 30 & 1000 & 0.030 & Bulk (ch0) & - \\
    Main & 30 & 2000 & 0.015 & Bulk (ch1) & - \\
    I/O  & -  & -    & 0.070 & Isochronous (ch0/1) & - \\
    I/O  & -  & -    & 0.080 & Bulk (ch0) & - \\
    I/O  & -  & -    & 0.070 & Bulk (ch1) & - \\
    \bottomrule
  \end{tabular}
\end{table*}

We run five configurations, summarized in Table~\ref{tab:throughput}. In each, the
device transmits 1024-byte packets on a single isochronous endpoint serviced once
per interval, and we vary only the burst size. The packet-size is based on the size limits of USB 3.x. SuperSpeed transactions group up to
bMaxBurst$+1$ packets into one burst, and with the burst multiplier held at zero
each interval carries a single such burst, which corresponds to the endpoint's entire
per-interval allowance. The five scenarios raise bMaxBurst from 0 to 15, which
doubles the packets delivered per interval from one to sixteen and, with them, the
theoretical throughput from 65.5\,Mbps to 1048.6\,Mbps. Each configuration runs for
180\,s, which is more than enough to ascertain throughput in steady state conditions. 

Across all five configurations the measured throughput lands within one to two
percent of the theoretical maximum, as the Achieved column of
Table~\ref{tab:throughput} shows: from 99.2\% at the smallest burst to 98.4\% at
the largest, where the link sustains 1032\,Mbps against a 1048.6\,Mbps ceiling. The
system therefore reaches at least 98\% of the theoretical limit at every operating
point, and saturates the isochronous schedule at roughly one gigabit per second.
The throughput is also stable within each run: the per-second minimum and maximum
differ by at most eight megabits per second.

\subsection{Latency Measurements}
\label{sec:eval:latency}

Sustained throughput does not by itself establish that a flow meets its timing requirements,
which for an isochronous stream is the binding constraint. We therefore measure
the per-packet interarrival latency of the isochronous flows while they share a
core with lower-priority bulk traffic. Each isochronous reader runs on a Quest Main VCPU,
whose period is set to half the endpoint's service interval. This ensures the reader samples
the flow at twice its arrival rate and the latest-only double-buffer of
Section~\ref{sec:design:host} never drops a packet between reads. We run two
isochronous flows of different periods, one serviced every 125\,$\mu$s and one
every 2\,ms, in each case alongside bulk traffic competing for the same core.

The VCPU parameters for all subsequent Quest experiments are shown in Table~\ref{tab:vcpu_params}, along with their corresponding flow specifications. Main VCPU budgets were established from measured latencies to process packet data. I/O VCPUs had dynamically calculated budgets and periods based on their specified utilization and the Main VCPU they served. These utilizations were sufficient to successfully support interrupt bottom half processing for the respective flow.  

\begin{figure*}[!htb]
  \centering
  \includegraphics[width=\textwidth]{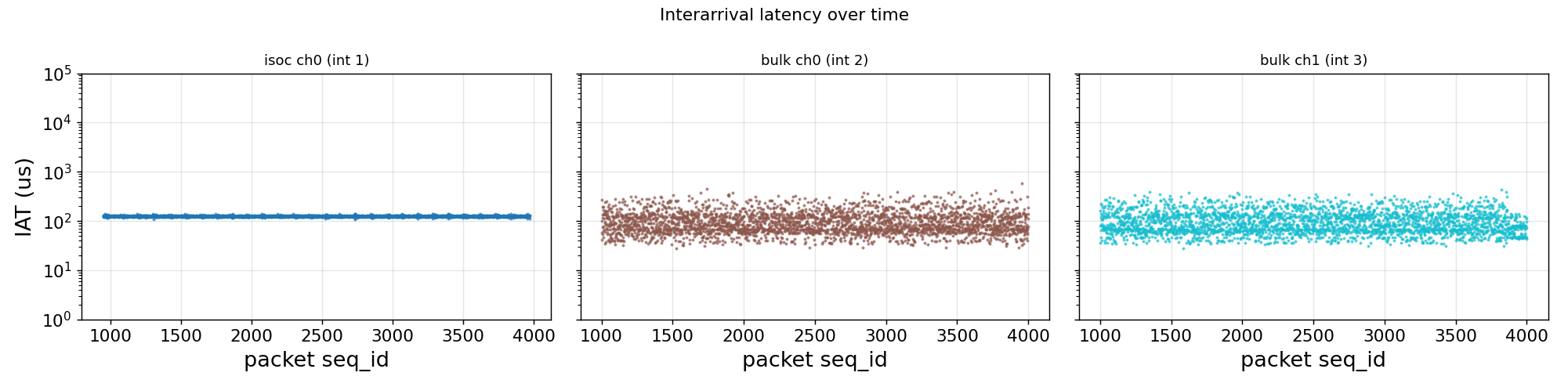}
  \caption{Per-packet interarrival latency (IAT) for a 125\,$\mu$s isochronous flow
    sharing a core with two bulk flows. The isochronous flow holds 
    tightly at its 125\,$\mu$s interval while the bulk flow shows significant jitter.
    % tight band at
    % its 125\,$\mu$s interval while the bulk flows show wide jitter.
    % , so it meets its
    % latency constraint despite the contention.
    }
  \label{fig:exp1-timeseries}
\end{figure*}

\begin{figure}[!htb]
  \centering
  \includegraphics[width=\columnwidth]{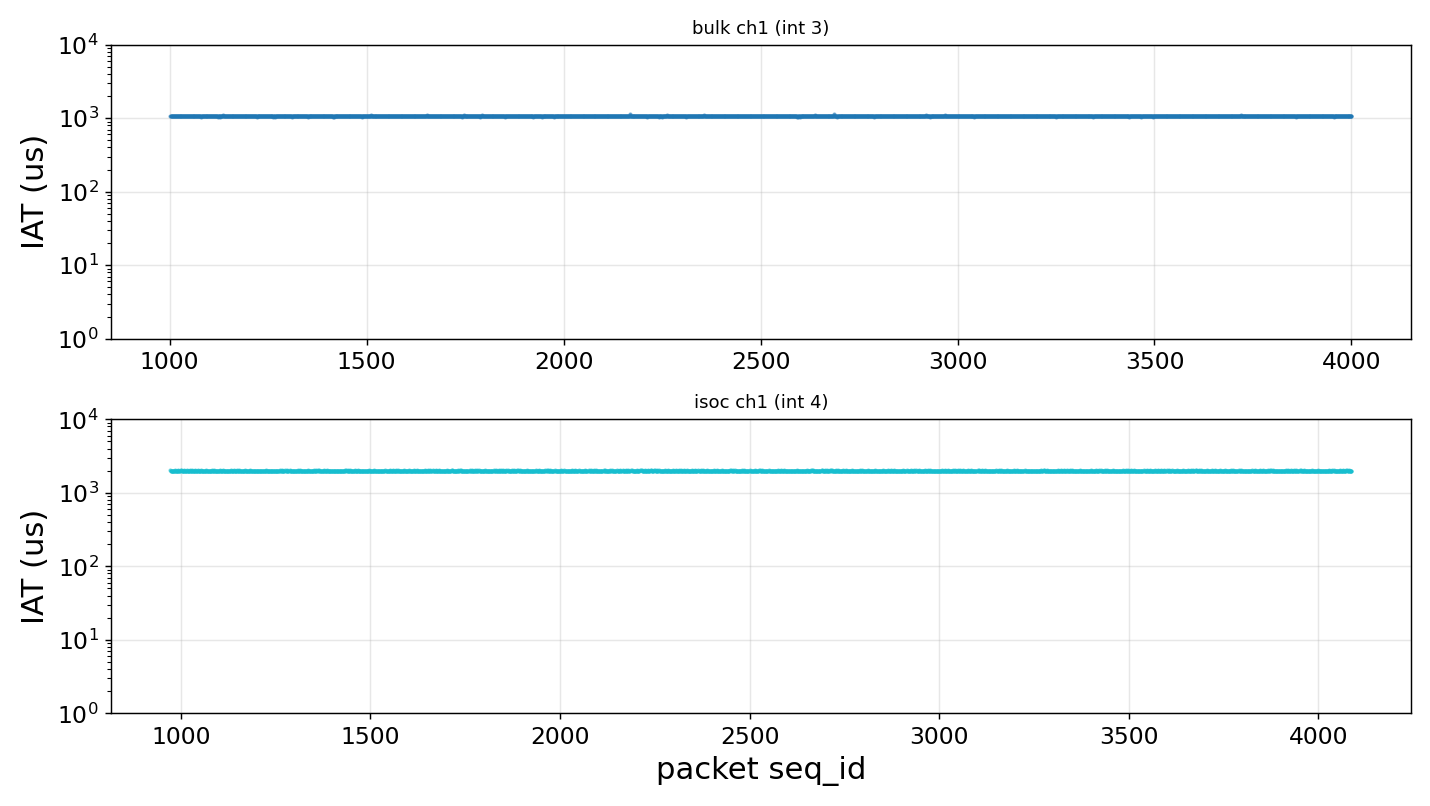}
  \caption{Per-packet IAT for a bInterval 2\,ms isochronous flow
    and a bulk flow
    % running under differentiated interrupters,
    with each stream assigned its own interrupter. The isochronous flow 
    % keeps 
    accurately maintains
    its IAT while sharing the core with bulk traffic.}
  \label{fig:1isoc1bulk}
\end{figure}

Both isochronous flows sustain their timing requirements. Figure~\ref{fig:exp1-timeseries} shows
the 125\,$\mu$s flow sharing its core with two bulk
streams: its interarrival latency stays in a tight band at the 125\,$\mu$s interval
while the bulk flows scatter across a far wider range. Figure~\ref{fig:1isoc1bulk} shows the 2\,ms flow alongside a
bulk flow, each routed through its own interrupter, and here both run steadily, the
isochronous flow at its 2\,ms interval and the bulk flow at its own cadence; with
sixteen times the slack of the 125\,$\mu$s case, the 2\,ms flow meets its constraint
with even greater margin. The isochronous flow maintains a tight interarrival latency because
differentiated service places its completion processing at the right priority: its
bottom half runs on a dedicated I/O VCPU (Section~\ref{sec:design:iovcpu}) whose
period tracks the flow's interval, so the kernel work that delivers each packet is
scheduled at the correct rate relative to competing bulk traffic, avoiding priority inversion.

% Wide time-series plot (~4:1) -> spans both columns (figure*).

\begin{figure}[!htb]
  \centering
  \includegraphics[width=\columnwidth]{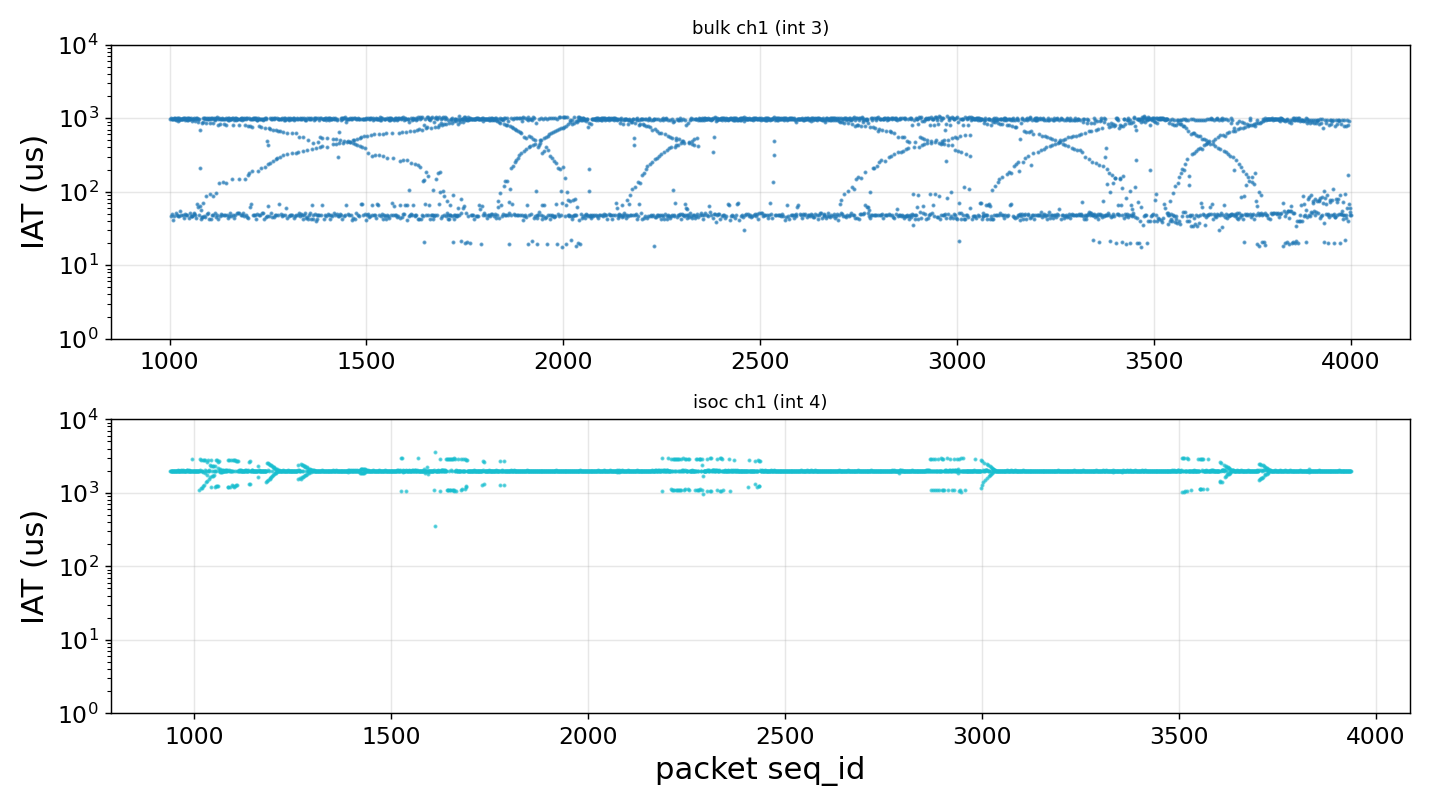}
  \caption{The same flows as Figure~\ref{fig:1isoc1bulk}, but with the isochronous
    and bulk streams sharing a single interrupter. Without differentiated service the
    isochronous flow's interarrival time shows periodic perturbations driven by the
    bulk traffic on the shared completion path.}
  \label{fig:single-interrupt}
\end{figure}

The value of differentiated service is clearest when it is taken away.
Figure~\ref{fig:single-interrupt} repeats this experiment with the same 2\,ms
isochronous flow and bulk flow, but routes both through a single shared interrupter,
so one bottom half now drains the completions of both. The isochronous flow no
longer holds a clean band: its interarrival time is punctuated by periodic bursts of
jitter, each driven by contention from the bulk traffic that now shares its
completion path, while the bulk flow itself spreads across a wide range. Assigning
each flow its own interrupter, as in Figure~\ref{fig:1isoc1bulk}, is what removes
these perturbations and lets the isochronous flow maintain its precise inter-arrival time.

\begin{figure}[!htb]
  \centering
  \includegraphics[width=\columnwidth]{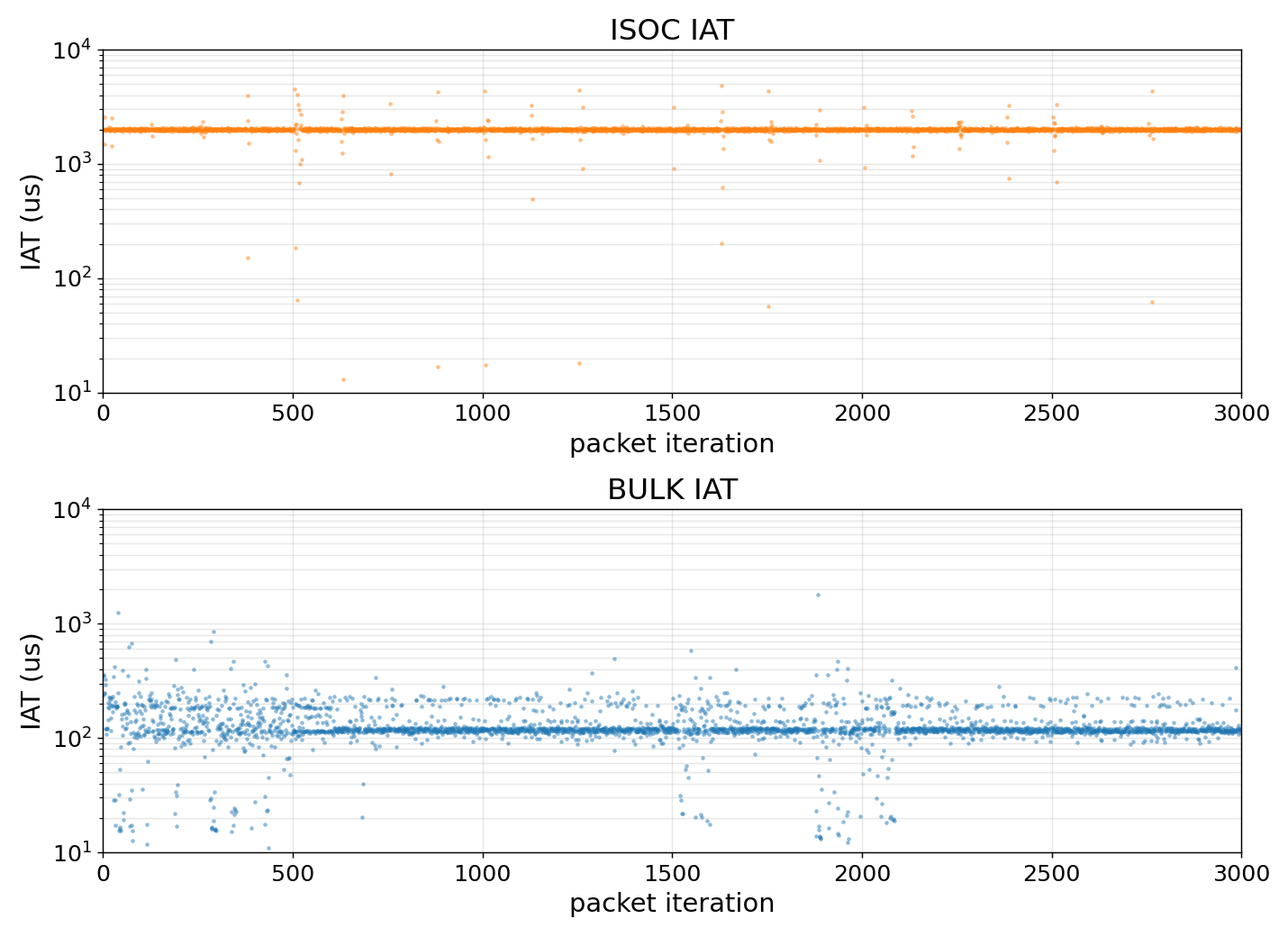}
  \caption{Linux-to-Linux reference: per-packet IAT for a 2\,ms
    isochronous flow and a bulk flow sharing a single interrupter, the only
    configuration Linux
    % currently 
    supports. 
    % The isochronous flow holds its 2\,ms
    % band but with sporadic outliers, and the bulk flow scatters about its mean.
    }
  \label{fig:linux-single}
\end{figure}

To place these results in context, we ran the same workload over a Linux-to-Linux
link as a reference. Commodity Linux currently supports only the single-interrupter
configuration: its stack offers no facility for routing separate flows to separate
completion paths, so the differentiated arrangement of Figure~\ref{fig:1isoc1bulk}
cannot be reproduced without modifying the I/O management subsystem. Figure~\ref{fig:linux-single} reports the Linux result
for the same 2\,ms isochronous flow and bulk flow sharing one interrupter. The
isochronous flow holds its 2\,ms band for most of the run but is broken by sporadic
outliers spanning more than two orders of magnitude, while the bulk flow experiences high latency variability about its mean. Lacking any means of isolating the isochronous completions
from the bulk traffic, Linux is unable to ensure tight temporal processing of isochronous traffic. There is no equivalent capability
to the per-flow differentiation that produces the uninterrupted isochronous timing
of Figure~\ref{fig:1isoc1bulk}.

\subsection{Dynamic Switching of Isochronous Flows}
\label{sec:eval:exp3}

A real deployment does not run a fixed set of flows for its entire lifetime. In a
vehicle, the rate a safety-critical flow requires depends on the driving situation.
A perception sensor such as a camera, LiDAR, or radar may stream at a modest rate
while the vehicle cruises on an open road, and then needs a much higher rate when an
obstacle appears or the vehicle enters a complex urban scene, where the control
loop must react on a tighter cycle. The bus must accommodate this change at runtime,
raising the flow's rate without disturbing its timing. This experiment shows that
our driver supports such a transition: an isochronous flow is switchable from one
service interval to another, mid-stream, while preserving the timing properties of
the active flow.

\begin{figure}[!htb]
  \centering
  \includegraphics[width=\columnwidth]{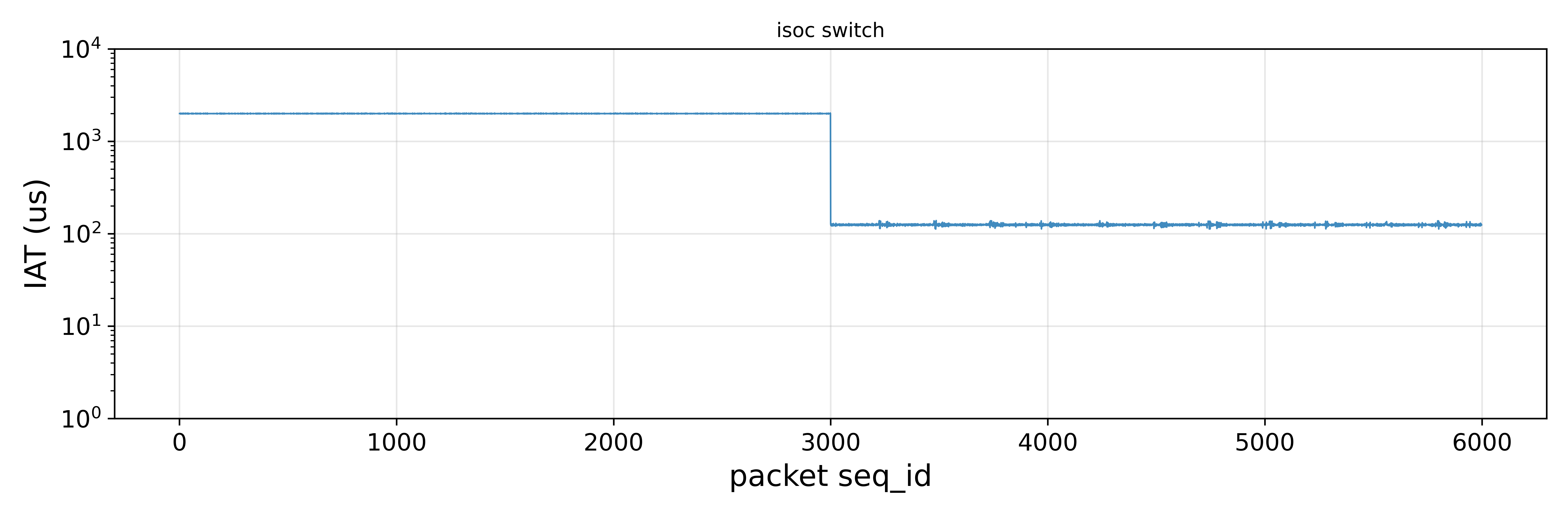}
  \caption{Switching an isochronous flow from a 2\,ms service interval (isoc1) to a
    125\,$\mu$s interval (isoc0) maintains expected IAT.
    %  midway through the run. The inter-arrival time holds
    % a tight band on both sides of the switch, so the flow keeps its isochronous
    % timing across the change in rate.
    }
  \label{fig:exp3-switch}
\end{figure}

Figure~\ref{fig:exp3-switch} reports a single isochronous channel whose service
interval is changed during the experimental run, while in the presence of the two bulk traffic flows as previously established using the VCPU paramters in Table~\ref{tab:vcpu_params}. We begin with a
2\,ms interval for the isochronous flow, and partway through the stream switch to a
125\,$\mu$s interval. The measured inter-arrival time tracks the change cleanly: it
holds a tight band at 2\,ms before the switch, and at 125\,$\mu$s after it, with no
loss of timing on either side of the transition. The flow therefore retains its
isochronous timing across a sixteen-fold change in rate, which is what a runtime
sensor reconfiguration requires.

% Single-column time-series (~3:1) -- stays within one column (regular figure,
% \columnwidth), per request. At one column it is short (~79pt tall); widen if needed.
% [H] forces the figure exactly here: the eval subsections have little text yet, so
% a normal float ([t]/[h]) gets deferred off the end and dropped. Revisit (back to
% [t]) once the eval prose fills in.
     % 4
\section{Related Work}
\label{sec:related}

% Six paragraph leads. Each paragraph closes with an explicit statement of how
% this work differs. Self-citations are third person (anonymous submission).
% Vocabulary and tone follow the Background primer (completion path, service
% class, flow, "left to chance").

\paragraph{Differentiated service in packet networks.}
Packet networks have offered differentiated service for decades. DiffServ marks
each IP packet with a class for per-hop forwarding~\cite{rfc2474, rfc2475}, and
IEEE 802.1Q VLAN tags carry a three-bit priority code point mapping to as many
as eight traffic classes~\cite{ieee8021q}. Differentiation also reaches inside
the sending host, where window-constrained schedulers bound each stream's losses
over a window of deadlines~\cite{west2004dwcs, mk-firm}. These mechanisms
classify, shape, or order traffic, but none governs the completion and interrupt
work a flow generates at the receiving host, nor extends to the peripheral bus
that carries the data its last few meters. A three-bit tag is also coarse: every
flow shares eight classes. The xHCI architecture this paper builds on provisions
as many as 1024 interrupters per controller~\cite{xhci-spec}, enough to give each
flow its own completion path rather than a place in one of eight queues
(Section~\ref{sec:bg:diffserv}).

\paragraph{Real-time buses.}
Buses built for control traffic guarantee timing through dedicated arbitration
or time slots. CAN arbitrates by fixed message priority~\cite{iso11898}, with
well-studied schedulability~\cite{davis2007can}. CAN~FD raises the data rate
but keeps CAN's arbitration model~\cite{hartwich2012canfd}, and FlexRay assigns
static TDMA slots alongside a dynamic segment~\cite{flexray30}. TSN brings
these disciplines to switched Ethernet through credit-based and time-aware
shaping over synchronized clocks~\cite{ieee8021q, ieee8021qbv, ieee8021as,
seol2021tsn}, whose gate scheduling is an active research
area~\cite{craciunas2016tsn}. All of these fabrics require dedicated wiring or
TSN-capable switches, and neither CAN~FD nor FlexRay exceeds 10~Mbit/s, whereas
USB offers Gbps rates and reserved-bandwidth iso\-chro\-nous transfers on ports
that ship with every commodity board. Njavro et al. compare USB against TSN for
combined networking and device I/O, finding USB paths achieve better average
and worst-case end-to-end latency even when TSN is granted large transfer windows~\cite{njavro2025real}. Results show that USB is able to achieve more than 
four times lower worst-case end-to-end latency compared to the best TSN case for 512-byte payloads under
Quest~\cite{njavro2025real}, when bridging networking and peripheral device buses. That study establishes USB as a viable CPS fabric
but treats the bus as one undifferentiated pipe; this paper supplies the
per-flow service guarantees it leaves open.

\paragraph{Host-to-host communication over peripheral links.}
Peripheral interconnects have been 
% pressed into service 
used
as host-to-host links
before. The USB communications device class carries Ethernet frames over a
device port~\cite{usb-cdc12}, and Linux implements it on both the host and
device sides. Linux's device side supports several industry-standard gadget drivers that emulate different Ethernet communication models over USB, including Ethernet Control Model (ECM), Ethernet Emulation Model (EEM), Network Control Model (NCM) and Microsoft's Remote Network Driver Interface Specification (RNDIS)~\cite{rndis}. Thunderbolt and USB4 define native host-to-host packets that Linux
exposes as a network interface~\cite{usb4-spec, linux-tbt-net}. PCIe
non-transparent bridges connect two PCIe hierarchies through shared memory
windows~\cite{linux-ntb}. West et al.\ repurpose the xHCI debug capability
(xDBC) as a full-duplex link between two USB~3.0 hosts at up to
5~Gbps~\cite{west2023real}. Each of these designs carries a single best-effort
stream. None offers isochronous service, and none distinguishes the flows
sharing the link. The xDCI gadget in this paper exposes every USB transfer type,
so one cable carries multiple flows, each under its own service class.

\paragraph{Quality of service for USB.}
Huang et al.\ retrofit QoS onto USB~1.1 and 2.0 by modifying endpoint
descriptors and scheduling transaction submission in the host controller
driver~\cite{huang2003usb, huang2004usb}. Later work gives USB transfers
temporal isolation on the host side of the bus. Missimer et al.\ bound
transaction delays in a real-time USB~2.0 (EHCI) host controller driver~\cite{missimer2013usb}.
Tuned Pipes compose end-to-end pipelines with throughput and delay guarantees
for USB~3.x devices~\cite{golchin2018tuned}, but offer no form of interrupt-processing differentiation on the end-hosts. Ruan et al.\ use xHCI interrupters to
give each device flow its own completion path with a per-flow
budget~\cite{usb-diff-services}. That work differentiates host-side device I/O
for bulk transfers involving USB~2 devices. This paper extends interrupter-based
differentiation across hosts through a device-side xDCI gadget, and covers
isochronous and bulk flows at USB~3 rates. Mainline Linux has also adopted
multiple interrupters. Intel added multi-interrupter support to Linux's xHCI
driver so that the completion events of offloaded audio endpoints post to a
secondary event ring that an audio DSP polls, letting the CPU sleep during
playback~\cite{linux-xhci-interrupters, linux-usb-audio-offload}. The mechanism
is shared with our approach, but the goal differs. Offloading removes the host CPU from the
completion path to save power, dedicating one extra interrupter to a single
device class, and providing no bandwidth or delay guarantees. The system
described here keeps the CPU in the path and disciplines it, with a budgeted
completion path for every flow under RTOS control.

\paragraph{Bandwidth-preserving scheduling.}
A bandwidth-pre\-serv\-ing server couples execution to a budget that replenishes
over a period. Whatever it runs cannot consume more than its reserved share of
the processor, no matter how much work arrives. The Deferrable Server and the
Sporadic Server established bandwidth preservation under fixed
priorities~\cite{strosnider1995deferrable, sprunt1989aperiodic,
bernat1999aperiodic}. Similarly, a family of dynamic-priority variants culminated in the
Constant Bandwidth Server~\cite{spuri1994tbs, ghazalie1995deadline,
deng1997open, abeni1998cbs}, and the POSIX realization of the
Sporadic Server~\cite{sporadic-server}. Processor capacity reserves made the
reservation a kernel abstraction in RT-Mach~\cite{mercer1994reserves}, and
resource kernels generalized it beyond the CPU~\cite{rajkumar1998resource}.

Later systems carried the budget into commodity settings, to account for
resource usage in servers~\cite{banga1999containers}, to compose
schedulers~\cite{regehr2001hls, shin2003periodic}, and to protect interactive
applications~\cite{yang2008redline}. Mainline Linux absorbed the idea, as CBS in
\texttt{SCHED\_DEADLINE}~\cite{lelli2016deadline} and as cgroup bandwidth
limits~\cite{turner2010cfs}. All of these approaches meter tasks. None of them meter
interrupt handling. \textsc{Preempt\_rt} makes interrupt handlers schedulable
threads~\cite{reghenzani2019preempt}, but a handler thread has no budget, so a
heavy interrupt load can starve lower-priority work for as long as the load
persists. Quest closes that gap on the host. Main VCPUs run tasks under Sporadic
Servers, and I/O VCPUs run interrupt bottom halves under PIBS, a
priority-inheritance bandwidth-preserving server~\cite{vcpu}. This approach
has been extended to mixed-criticality I/O scheduling~\cite{mixed-crit-io}. Binding each I/O VCPU to
a separate interrupter enables an interrupt processing reservation for a single device flow, rather than one reservation for all flows~\cite{usb-diff-services}. This paper extends
that binding to flows that cross the bus between hosts as well as peripheral devices, for both bulk and isochronous endpoints.

{\emergencystretch=2em
\paragraph{Early interrupt demultiplexing.}
An interrupt on a shared completion path does not identify the flow it serves;
software learns that only by inspecting completion state after the interrupt
has claimed the CPU, so attribution consumes processor cycles outside the flow's budget. A
saturated host can be consumed by packet interrupts, without ever finishing other work, leading to receive livelock~\cite{mogul1997livelock}. Lazy
Receiver Processing demultiplexes packets early and charges protocol processing
to the receiving process~\cite{druschel1996lrp}. This restores accounting but
not isolation. Demultiplexing still preempts whatever task holds the CPU. Zhang
and West account deferred interrupt work at the priority of the process
awaiting the data~\cite{zhang2006interrupt}. Parmer and West make interrupt
execution schedulable end to end in Composite~\cite{parmer2008interrupt}.
Leyva-del-Foyo et al.\ unify interrupt and task scheduling under one priority
space, though priorities are fixed per source and
unmetered~\cite{leyva2006interrupt}, while Lewandowski et al.\ characterize a
network driver's CPU demand to account for its
interference~\cite{lewandowski2007driver}. In each of these works, the classification step is performed by software. USB interrupters move this into the host controller.
Completions arrive pre-sorted by flow, each on its own event ring, drained by
a corresponding I/O VCPU. This paper couples that hardware demultiplexing
with bandwidth-preserving servers at both ends of the bus.\par}
        % 5  positions the "first" claim
\section{Conclusion and Future Work}
\label{sec:conclusion}

% Merged section (Modeling removed; the old "Discussion and Limitations" is folded
% in here). TODO:
%   - Recap the contribution and the headline result.
%   - Limitations / dependencies: generality beyond Quest and portability of the
%     IO/VCPU co-design; hardware / USB-controller dependence; scaling past two
%     hosts and topology assumptions; security.
%   - Future work: scaling, topologies, device-I/O applications, integration with
%     safety / availability mechanisms.

This paper introduces a unified USB communication approach, supporting
host-to-host and host-device communication. The work specifically
addresses the use of an xDCI gadget driver for USB 3.x communication
with configurable numbers of endpoints and types, including bulk and
isochronous pipes. These are associated with separate interrupters,
which are tied to time-budgeted bottom half threads in Quest. These
bottom halves are scheduled according to the interrupter associated
with the transfer completion, and they execute on schedulable I/O
VCPUs that have a guaranteed budget and replenishment period. This
ensures interrupts are correctly prioritized, with their handlers
guaranteed a processor capacity reserve according to rate-monotonic
scheduling theory.

Quest is the first system to combine time-budgeted task and interrupt
processing with communication over USB, supporting differentiated
service across USB 3.x links. Future work will investigate more
complex bus topologies 
% besides point-to-point communication, 
using
combinations of xDCI, xHCI and xDBC endpoints. USB has the potential
to be a replacement control area networking bus for distributed
embedded systems found on emerging robotics and automotive platforms.
     % 6  Conclusion and Future Work (incl. limitations)

%% ---------------------------------------------------------------------------
%% Bibliography
%% ---------------------------------------------------------------------------
\bibliographystyle{ACM-Reference-Format}
\bibliography{references}

\end{document}